\documentclass[a4paper,11pt]{article}
\usepackage{jinstpub} 
\usepackage{lineno}
\usepackage{xcolor}
\usepackage{multicol}
\usepackage{tikz}
\usepackage{subcaption}
\usepackage{gensymb}
\usetikzlibrary{shapes.geometric, arrows.meta, positioning}
\hypersetup{
  colorlinks=true,
  linkcolor=blue,
  citecolor=blue,
  urlcolor=blue
}

\title{\boldmath Investigation of Thick-GEM detectors fabricated in India for muography application}

\author[a]{Saikat Ghosh}
\author[b,1]{Promita Roy \note{Corresponding author}}
\author[c]{Subhendu Das}
\author[a]{Shubhabrata Dutta}
\author[a]{Nilanjan Biswas}
\author[d]{Supratik Mukhopadhyay}
\author[a]{Nayana Majumdar}
\affiliation[a]{Saha Institute of Nuclear Physics, A CI of Homi Bhabha National Institute,\\
1/AF Bidhannagar, Sector-I, Kolkata 700064, India}
\affiliation[b]{Centre for Neutrino Physics, Virginia Polytechnic Institute and State University,\\
Blacksburg 24061, VA, United States}
\affiliation[c]{Depatment of Physics, Government General Degree College,\\ 
Chapra, Nadia 741123, West Bengal, India}
\affiliation[d]{Research Wing, Naihati Prolife,\\
Naihati 743165, West Bengal, India}

\emailAdd{promita@vt.edu}

\abstract{Muography, commonly known as muon tomography, is a passive, non-destructive imaging technique that utilizes naturally occurring cosmic-ray muons to visualize the internal density structures of large, static, or inaccessible objects. In course of developing a small prototype muography system for material identification that relies upon the multiple Coulomb scattering of muons in matter, we explored the possible use of Thick-GEM detector as muon tracking device. It is a 5-20 fold scaled-up version of traditional GEM technology, that has become increasingly popular in recent years, owing to its mechanical robustness, cost-effective production, and excellent position sensing capabilities. A few prototypes of this detector of dimension $40\,\mathrm{mm} \times 48\,\mathrm{mm}$ with variation in other design parameters, were manufactured from a local industry. Subsequent to conditioning, detailed characterization of the detectors was performed to validate their suitability in muography applications.  
To identify the optimal operating region, gain variation was studied under various voltage configurations for both single and double-stage configurations. Experimental measurement of muon detection efficiency across the entire operating range yielded a maximum efficiency of 99.5\% in both cases. 
Using a collimated Fe$^{55}$-source, the best spatial resolution was determined to be 30 $\mu$m for both single and double-stage operation.}

\keywords{Gaseous detectors, Gaseous imaging and tracking detectors, Micropattern gaseous detectors (MSGC, GEM, THGEM, RETHGEM, MHSP, MICROPIC, MICROMEGAS, InGrid, etc), Particle tracking detectors (Gaseous detectors). }

\begin{document}
\maketitle
\flushbottom

\section{Introduction}
\label{sec:intro}

Cosmic-ray muons are naturally produced in the atmosphere and arrive at sea level with a broad energy spectrum and an angular distribution that make them available “for free” everywhere on the Earth. Because muons are minimum-ionizing and highly penetrating, they can traverse tens to hundreds of meters of a medium (depending on energy and material) while still remaining detectable. Muon tomography, often termed as muography, exploits this property to infer on internal density variations in large or shielded structures without using artificial radiation sources.

The historical origin of muon tomography or muography is often traced back to early absorption measurements for overburden studies and, most notably, to the pioneering work by Alvarez and collaborators who used cosmic muon absorption to search for hidden chambers in the Pyramid of Khafre~\cite{Alvarez1970}. That experiment demonstrated that directional muon flux measurements, when combined with tracking, could constrain density variations inside monumental structures and established a foundation for modern transmission muography.

Contemporary muography broadly employs two complementary modalities. In transmission (attenuation) muography, one measures the directional reduction of muon flux through a target relative to open-sky expectations and reconstructs the integrated density (“opacity”) along each line of sight. This approach is particularly powerful for large-scale targets such as volcanoes and geological structures, where path lengths can be very large and the signal is primarily a deficit in the transmitted muon rate~\cite{Morishima2017,LoPresti2020}. Modern reviews discuss the rapid growth of this field and highlight the breadth of present-day applications~\cite{Lechmann2021,Yang2018}.

A second major class is scattering muography, which uses the angular deflection of muons caused by multiple-Coulomb-scattering (MCS) as they traverse matter~\cite{Schultz2004}. Since scattering depends on radiation length and is more pronounced for dense/high-Z materials, scattering muography provides enhanced sensitivity to compact high-Z objects even when they are shielded by lower-Z materials, making it attractive for security screening and safeguards~\cite{Priedhorsky2003,Barnes2023}. Foundational work established reconstruction approaches and demonstrated material discrimination using measured muon scattering distributions.

The performance and practicality of a muography setup are strongly influenced by the choice of detector technology. Typical requirements include robust detector design, large geometric acceptance, good spatial resolution to enable accurate track reconstruction, high detection efficiency for minimum-ionizing particles, and stable long-term operation~\cite{Giammanco2025}. Since many muography deployments involve extended data-taking periods in non-laboratory environments, detector robustness, scalability, and cost-effectiveness are often as critical as intrinsic performance~\cite{Bonechi2020}. Consequently, a wide range of detector technologies were explored for muography, including plastic scintillators~\cite{Ambrosino2014, Bajou2023}, drift tubes and drift chambers~\cite{Kume2016, Yang2023}, resistive plate chambers (RPCs)~\cite{Ambrosino2015, Wang2015MRPC, Sarmento2024}, nuclear emulsions~\cite{Bozza2017}, and micro-pattern gaseous detectors (MPGDs)~\cite{Bouteille2016, Wang2025, Gnanvo2011, Biglietti2016}. 

Among these, MPGDs have attracted increasing attention due to their fine granularity, flexible detector geometry, and compatibility with modern readout electronics. Gas Electron Multipliers (GEMs) and Micromegas have demonstrated excellent tracking performance in high-energy physics and related applications, motivating their consideration for muography systems. Within this family, the \textit{Thick Gas Electron Multiplier} (THGEM) stands out for its robustness, mechanical simplicity, and scalability to large areas. Initially inspired by the concept of GEM, introduced by F. Sauli in 1997~\cite{Sauli1997}, THGEMs have been developed to enhance mechanical stability and simplify manufacturing, without compromising performance~\cite{Chechik2004}. THGEMs are fabricated by mechanically drilling regular hole patterns into thick insulating substrates clad with conductive electrodes, typically using standard printed-circuit-board (PCB) techniques~\cite{Breskin2009, Roy2023}. Compared to conventional thin GEM foils, THGEMs offer enhanced mechanical robustness, simplified handling, and greater tolerance to imperfections, all of which are advantageous for detectors intended for large-scale or field-deployable systems~\cite{Breskin2009}. At the same time THGEMs can achieve gains as high as $10^4$--$10^5$ depending on the gas mixture, pressure, and operational conditions~\cite{Shalem2006PartI, Alexeev2015}, while multi-stage cascaded THGEM systems can achieve even higher effective gains with improved discharge stability~\cite{Cortesi2009,Arazi2010}.
Systematic studies have been carried out to investigate the dependence of gain, stability, and discharge probability on geometrical parameters such as hole diameter, pitch, substrate thickness, and rim etching~\cite{Roy2023, Shalem2006PartI}. The effects of charging-up~\cite{Pitt2018}, gain uniformity~\cite{Li2021}, and long-term stability under sustained irradiation have also been examined, providing important insights into operational behavior relevant for prolonged data-taking scenarios. THGEM operation has been demonstrated in a variety of gas mixtures and pressures, highlighting the flexibility of the technology and its adaptability to different experimental conditions~\cite{Alon2008, Dai2026}. 

Spatial resolution is a key performance metric for muography detectors. Understanding how THGEM design and readout choices influence this metric has been the subject of significant research. Early studies of THGEM spatial performance primarily focused on charge transport, avalanche development, and signal formation, establishing that THGEM-based detectors with optimized hole geometry and segmented readout can achieve sub-millimeter spatial resolution under laboratory conditions~\cite{Bressler2023THGEM}. For example, cosmic-ray tests of a mini-drift THGEM chamber reported a spatial resolution of about $220\,\mu$m, demonstrating that appropriately configured THGEM-based detectors can provide precise tracking performance~\cite{Yang2015MiniDriftTHGEM}. In another study, Cortesi \textit{et al.} investigated a THGEM-based imaging detector and reported a spatial resolution of about $0.7$ mm FWHM with a resistive-anode 2D readout, further demonstrating the imaging capability of THGEM-based detectors under laboratory conditions~\cite{Cortesi2007Imaging}. More recently, Zhao \textit{et al.} developed and studied THGEMs with reduced hole and pitch sizes of few 100s of micron, which were found to have sufficiently good position resolution (close to $75\,\mu$m ) for muon tracking and imaging~\cite{Zhao2017HighResTHGEM}.

In order to realize a muography system based on THGEM technology, the development of large-area detector foils is a critical requirement. As an initial and essential step toward this goal, we designed and fabricated several small-scale THGEM prototypes using foils with varied design parameters, in collaboration with a few local industries as a continuation of our earlier study~\cite{Roy2023}. 
The prototypes built with the fabricated variants of the THGEM foil underwent rigorous characterization by studying their gas gain, following the application of multiple conditioning techniques to the foils. On the basis of performance, one of the prototypes was opted for further investigations and subjected to extensive performance evaluation with two different gas mixtures. The key requisites for muon detectors in a muography setup, the muon detection efficiency and the position resolution of the prototype were investigated using both the single and the double-stage configuration to examine the suitability of multi-stage operation.

The present article is organized in the following way to report all these activities and relevant findings. The Section~\ref{THGEM_proto} discusses about the locally fabricated THGEM foils, and their conditioning. It is followed by discussions on the performance evaluation of the prototypes built with THGEM foils in single and double-stage configuration using two different Argon-based gas mixtures. The measurement of muon detection efficiency and the position resolution of the THGEM prototype built with the foils are discussed in Section~\ref{Efficiency} and Section~\ref{Resolution}, respectively. The article ends with Section~\ref{Conclusion} where conclusive remarks are made based on the findings of this study.

\section{THGEM prototypes}
\label{THGEM_proto}
The THGEM foils were intentionally designed and produced with variations in substrate material, thickness, and key geometrical parameters, to allow a systematic investigation of how such choices influence the detector performance, such as the gas gain, stability, discharge probability, etc. Upon procurement of the produces from the local manufacturers, the THGEM foils underwent multiple conditioning techniques. These conditioning procedures play a crucial role in stabilizing detector operation by removing trapped moisture, cleaning residues and reducing surface asperities which in turn help to suppress micro-discharges and improve long-term gain stability.
Following their conditioning, THGEM detector prototypes were made by assembling different types of foils with suitable arrangement of drift and anode readout planes. On the basis of measured gains over a range of electric-field configurations, the prototype with the best-performing foil was subjected to further investigations. The THGEM prototype made with the specific foil in single as well as double-stage configuration, was characterized extensively by studying its gas gain and energy resolution using two different Ar-based gas mixtures. In the following sub-sections, these aspects are reported in details.

\subsection{Design and fabrication}
\label{Design}
\begin{figure}[htbp]
\centering
\includegraphics[width=0.35\textwidth]{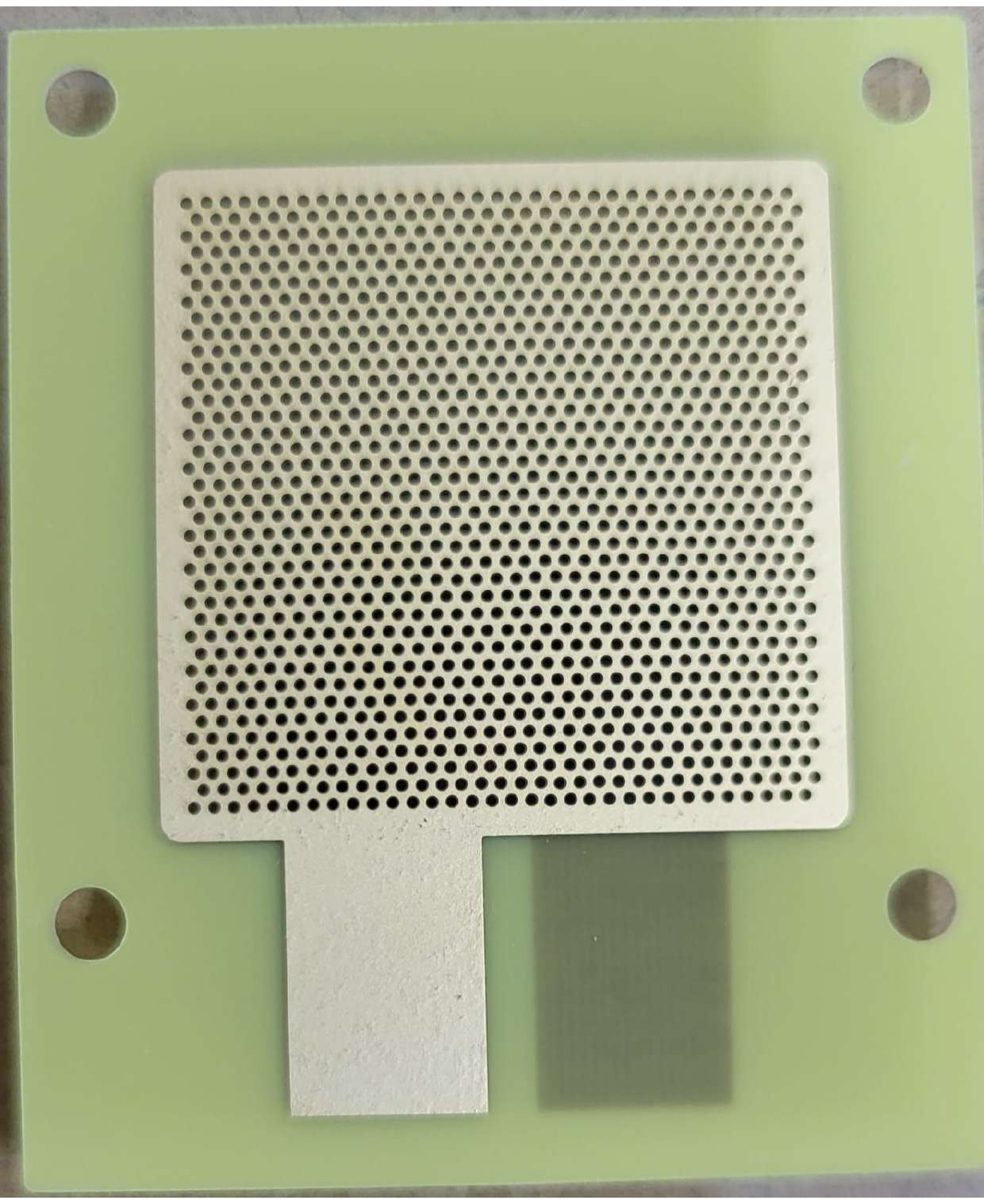}
\caption{One of the locally fabricated THGEM foils (THGEM-B)
\label{fig:foil image}}
\end{figure}

The THGEM foils, fabricated for this study, were of dimension $40\,\mathrm{mm} \times 48\,\mathrm{mm}$, with an active area $28\,\mathrm{mm} \times 28\,\mathrm{mm}$, and different thickness: $269$, $359$ and $400\ \mathrm{\mu m}$. The substrate material, selected for the insulating layer of the foil, was double-sided FR4 PCB of two different variants, namely, Isola IS410 and ITEQ IT-180A. The material specifications are provided in Appendix~\ref{mat_table}. The copper thickness used in each case is $35\ \mathrm{\mu m}$. In the active area, non-plated through holes (NPTH) of diameters $400$ and $500\,\mu\mathrm{m}$, were produced in a hexagonal pattern with pitch $800\ \mu\mathrm{m}$ and $1\ \mathrm{mm}$, respectively, by industrial drilling of the substrate with a tolerance limit $50\,\mu\mathrm{m}$. To mitigate the occurrence of electrical discharges caused by high voltages, applied to the conducting surfaces on either sides of the THGEM foil, rims of widths $50$ and $100\,\mu\mathrm{m}$ were etched around each hole through standard chemical processes. A final surface finish of electroless nickel immersion gold (ENIG), consisting of a nickel layer of $4.58\,\mu\mathrm{m}$ thickness, followed by a $0.0678\,\mu\mathrm{m}$ thick gold layer, was applied on the copper coated surfaces of the FR4 PCB. The material and design parameters of the THGEM foils (namely, A, B, and C), are tabulated in Table~\ref{tab:Design_param}. A sample of the local produces of the THGEM (THGEM-B) foils is depicted in figure~\ref{fig:foil image}.
\begin{table}[htbp]
\centering
\caption{Material and design parameters of THGEM foils\label{tab:Design_param}}
\smallskip
\begin{tabular}{|c|c|c|c|c|c|}
\hline
Name&Material&Thickness&Hole diameter&Rim width&Pitch\\
& &($\mathrm{mm}$)&($\mathrm{mm}$)&($\mathrm{mm}$)&($\mathrm{mm}$)\\
\hline
THGEM-A & Isola IS410 & 0.359 & 0.5 & 0.1 & 1.0\\
THGEM-B & ITEQ IT-180A & 0.269 & 0.4 & 0.05 & 0.8\\
THGEM-C & ITEQ IT-180A & 0.4 & 0.4 & 0.05 & 0.8\\
\hline
\end{tabular}
\end{table}

\subsection{Cleaning and conditioning}
\label{Condition}
 
\begin{figure}[htbp]
\centering
\includegraphics[width=0.8\textwidth]{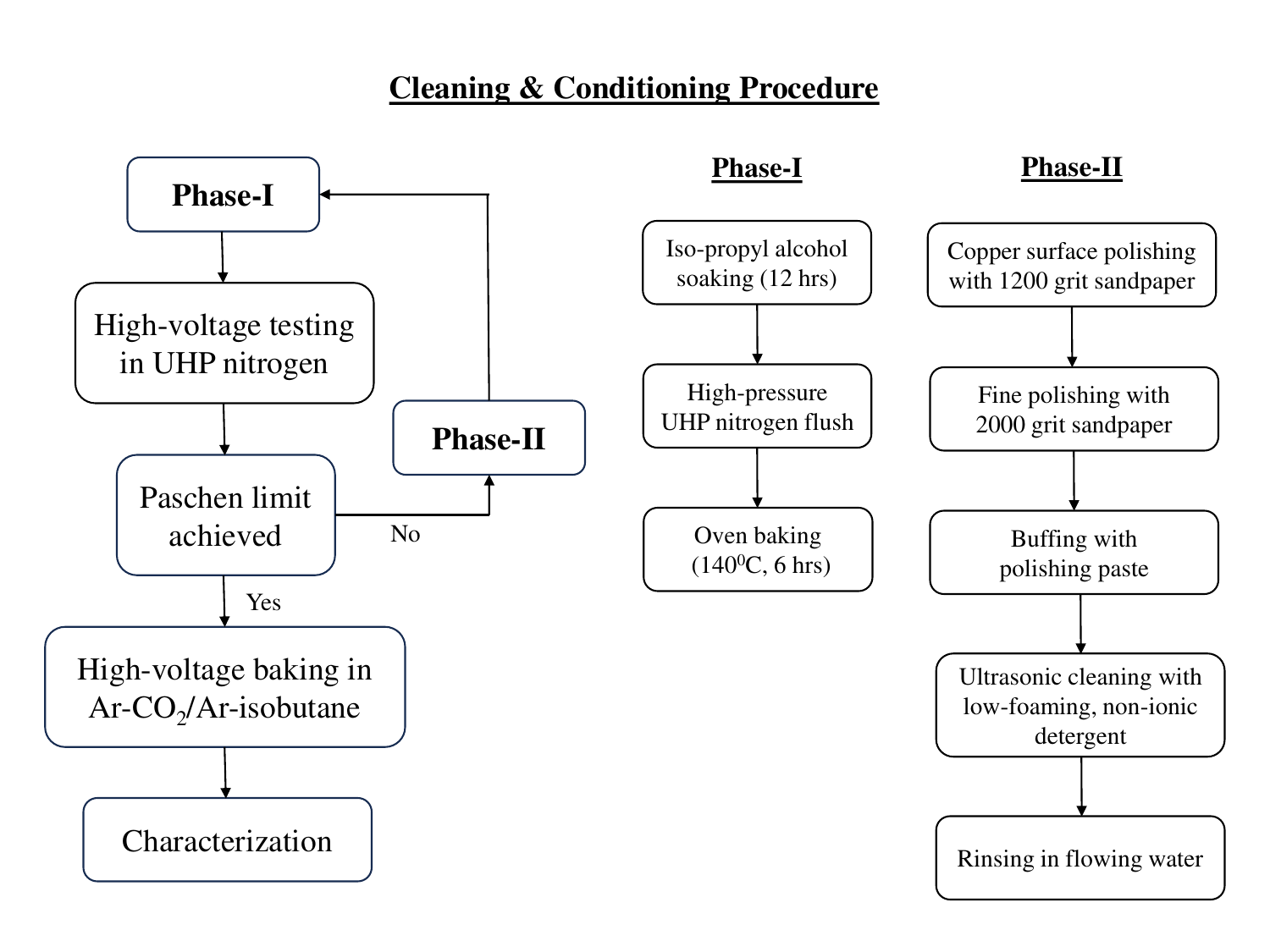}
\caption{Flowchart of the THGEM foil cleaning and conditioning procedure
\label{fig:procedure} }
\end{figure}

A series of cleaning and conditioning procedures was followed to prepare the foils for application of high voltage across the conducting surfaces with an objective of attaining the Paschen limit in ultra-high purity (UHP) nitrogen. As illustrated in the flowchart in figure \ref{fig:procedure}, the procedure was accomplished in two phases (phase-I, phase-II). The phase-I started with soaking the foils in pure isopropyl alcohol for almost $12\,\mathrm{hrs}$. The foils were then flushed with high-pressure, UHP nitrogen jet, followed by baking in oven for $6-8\,\mathrm{hrs}$ at $140\,\mathrm{\degree C}$. They were then subjected to high-voltage testing in a gas tight chamber with continuous UHP nitrogen flow by applying and gradually raising a potential difference, $\Delta V_{THGEM}$, across the THGEM foil, until discharge was observed. Initially at this stage, the foils THGEM-A and THGEM-C exhibited breakdown at $\Delta V_{THGEM}$, below the expected Paschen limit (< 80\%), while THGEM-B could reach upto 96\% of the limit.
 
To reduce the discharge probability and achieve the Paschen limit, a systematic polishing and cleaning procedure for those two THGEM foils (A, and C) was adopted in phase-II. These steps included polishing of the conducting surfaces of the foils with $1200\,\mathrm{grit}$ sandpaper initially to remove coarse surface asperities, followed by fine polishing with $2000\,\mathrm{grit}$ sandpaper to achieve improved smoothness. Subsequently, buffing was performed using polishing paste to smoothen the copper edges and reduce sharp features around the hole rims. Following the mechanical treatment, the THGEM foils were cleaned in an ultrasonic bath containing a low-foaming, non-ionic detergent for $1\,\mathrm{hr}$ to remove residual dust, particularly from within the holes. The foils were then thoroughly rinsed under flowing water. The polishing and cleaning procedures applied to the THGEMs were initially inspired by the methodology reported in~\cite{Alexeev2014}, and subsequently refined through the implementation of modified approaches developed in this work. Finally, the foils were again subjected to the phase-I cleaning procedure to make them ready for the high-voltage testing.

\begin{figure}[htbp]
\centering
   \subfloat[ \label{fig:microscope_img}]{%
    \includegraphics[width=0.5\textwidth]{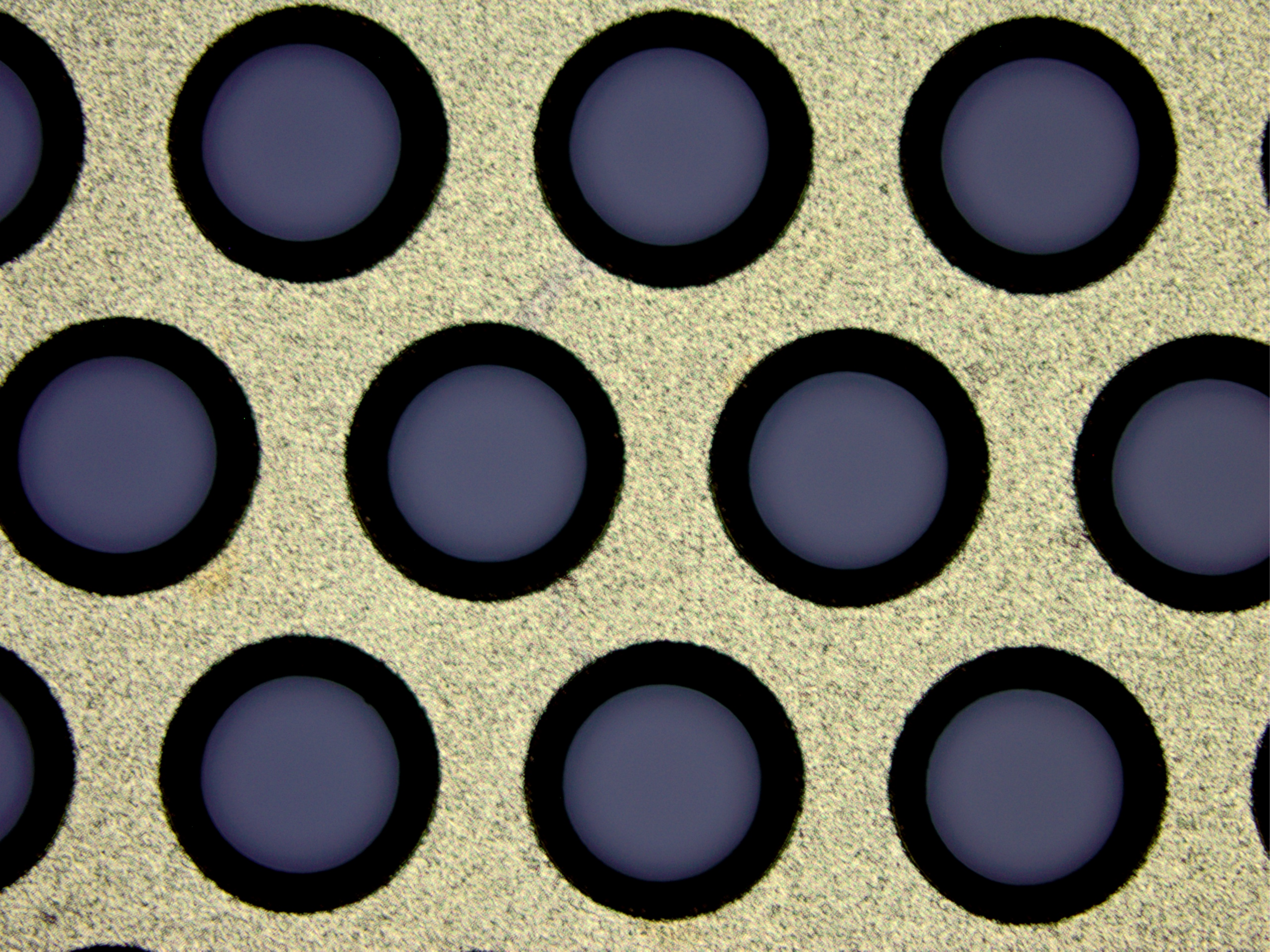}}
  \hfill
   \subfloat[\label{fig:microscope_img_polished}]{%
    \includegraphics[width=0.5\textwidth]{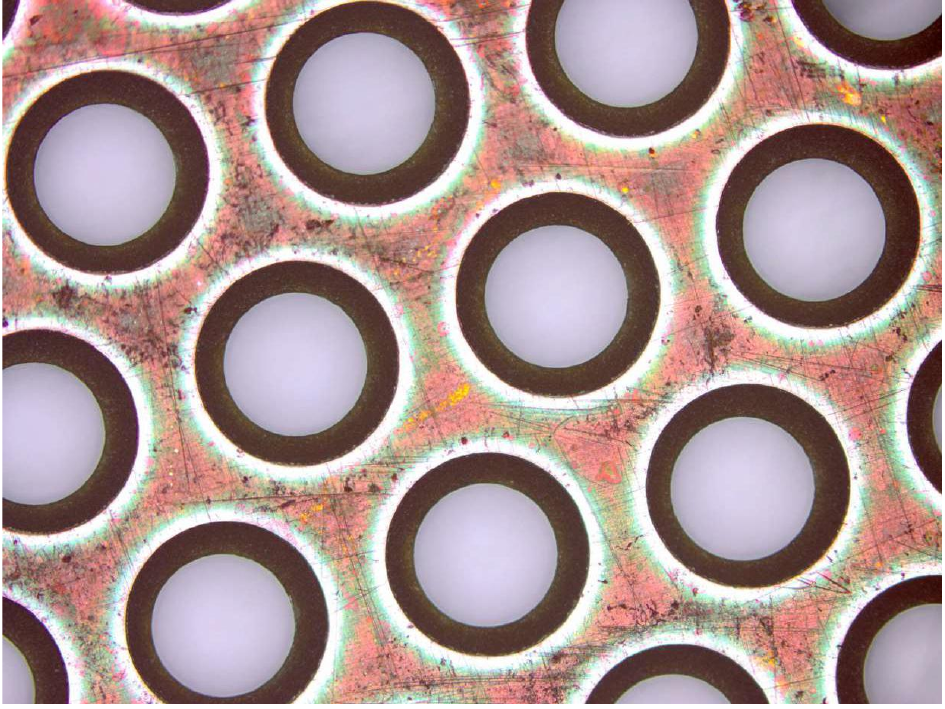}}
\caption[]{\label{fig:microscope} THGEM foil after (a) phase-I (5X magnification), and (b) phase-II (5X magnification) conditioning}
\end{figure}

A comparison of microscopic images after carrying out the phase-I and the phase-II followed by repeating phase-I, is shown in figure~\ref{fig:microscope}. The ENIG surface-finish, originally applied by the manufacturer, was removed during the polishing of the foils in phase-II. As a result, some surface oxidation occurred on the bare copper electrodes during high-temperature baking due to exposure to the atmosphere which is evident from the figure~\ref{fig:microscope}. However, this did not affect the overall performance of the detectors, rather improved the maximum limit of the applicable $\Delta V_{THGEM}$ in UHP nitrogen. The $\Delta V_{THGEM}$ limits before and after the phase-II treatment for the THGEM foils (A and C) along with THGEM-B which did not require the additional phase-II conditioning, are summarized in Table~\ref{tab:Paschen_limit}, together with the corresponding Paschen limits, calculated using the empirical equations reported in~\cite{Husain1982} and considering only the thickness of the foils as distance between the electrodes.

\begin{table}[htbp]
\centering
\caption{High-voltage breakdown limits of THGEM foils in UHP nitrogen\label{tab:Paschen_limit}}
\smallskip
\begin{tabular}{|c|c|c|c|}
\hline
Name&$\Delta V_{THGEM}\,\mathrm{(V)}$&$\Delta V_{THGEM}\,\mathrm{(V)}$&Paschen limit $\mathrm{(V)}$\\
 & (after phase-I) & (after phase-II) & \cite{Husain1982}\\
\hline
THGEM-A & 1800 & 2600 & 2310\\
THGEM-B & 1950 & -- & 2023\\
THGEM-C & 1850 & 2350 & 2489\\
\hline
\end{tabular}
\end{table}

\begin{figure}[htbp]
\centering
\includegraphics[width=0.6\textwidth]{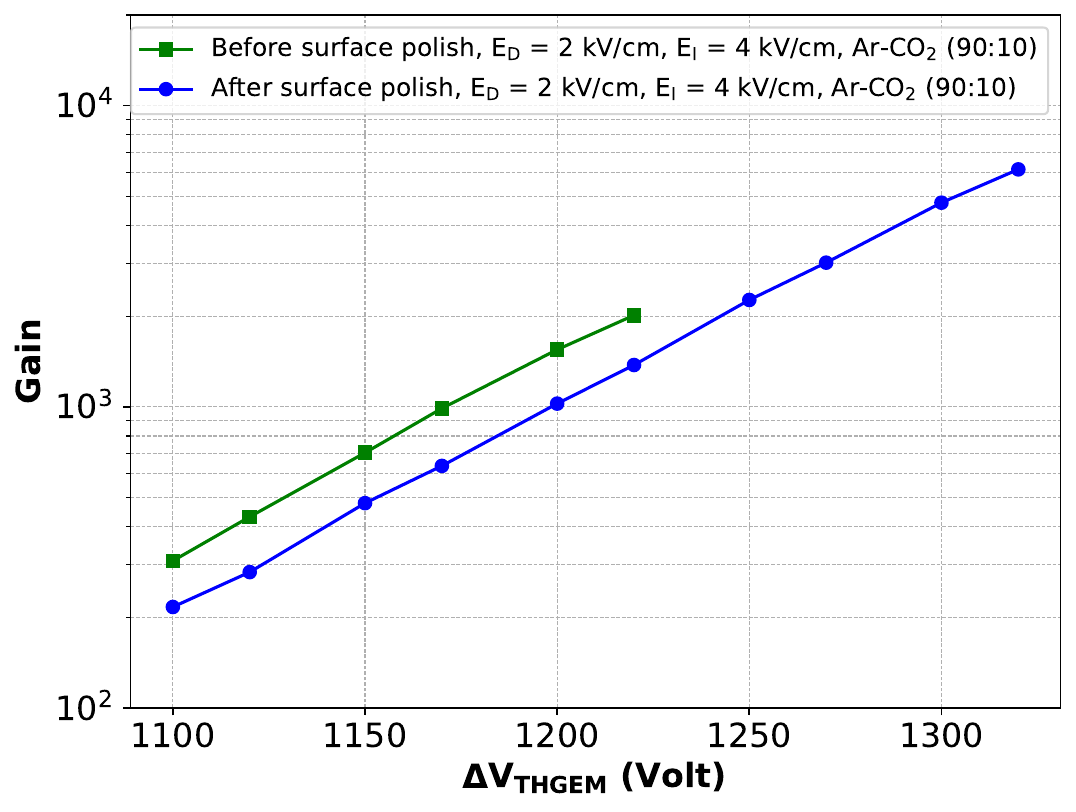}
\caption{Comparison of performance of THGEM prototype detector (made with THGEM-C) before and after phase-II conditioning \label{fig:polishing_effect}}
\end{figure}

The overall improvement in the performance of the THGEM foils due to phase-II conditioning can be observed from figure \ref{fig:polishing_effect} where the gains measured in Ar-CO$_2$ (90:10) before and after the phase-II conditioning are plotted for THGEM-C, as an example. It shows that the gain reduced at a given $\Delta V_{THGEM}$ due to phase-II conditioning, nevertheless, the working regime could be extended beyond $\Delta V_{THGEM}=1220\,\mathrm{V}$ which was the limit in pre-polishing stage. As a result, three-fold enhancement in the gain value could be achieved. The details about the process of gain characterization is discussed in the next section.

\section{Characterization studies}
\label{Char_study}
Following systematic conditioning of the THGEM foils, the performance of the THGEM prototype detectors, made with the foils in single-stage configuration, was studied by measuring their gas gain over a range of applied $\Delta V_{THGEM}$. Further detailed investigation on the performance of the detector with the best-performing foil (THGEM-B), was carried out in both of the single and double-stage operation. Since the choice of operating gas significantly impacts charge amplification, operational stability, and spatial resolution, the detector performance was studied using two different Argon-based gas mixtures at atmospheric pressure, enabling a comparative assessment for optimized operation.

\subsection{Single-stage set-up}
\label{STHGEM}
\begin{figure}[htbp]
\centering
\includegraphics[width=0.7\textwidth]{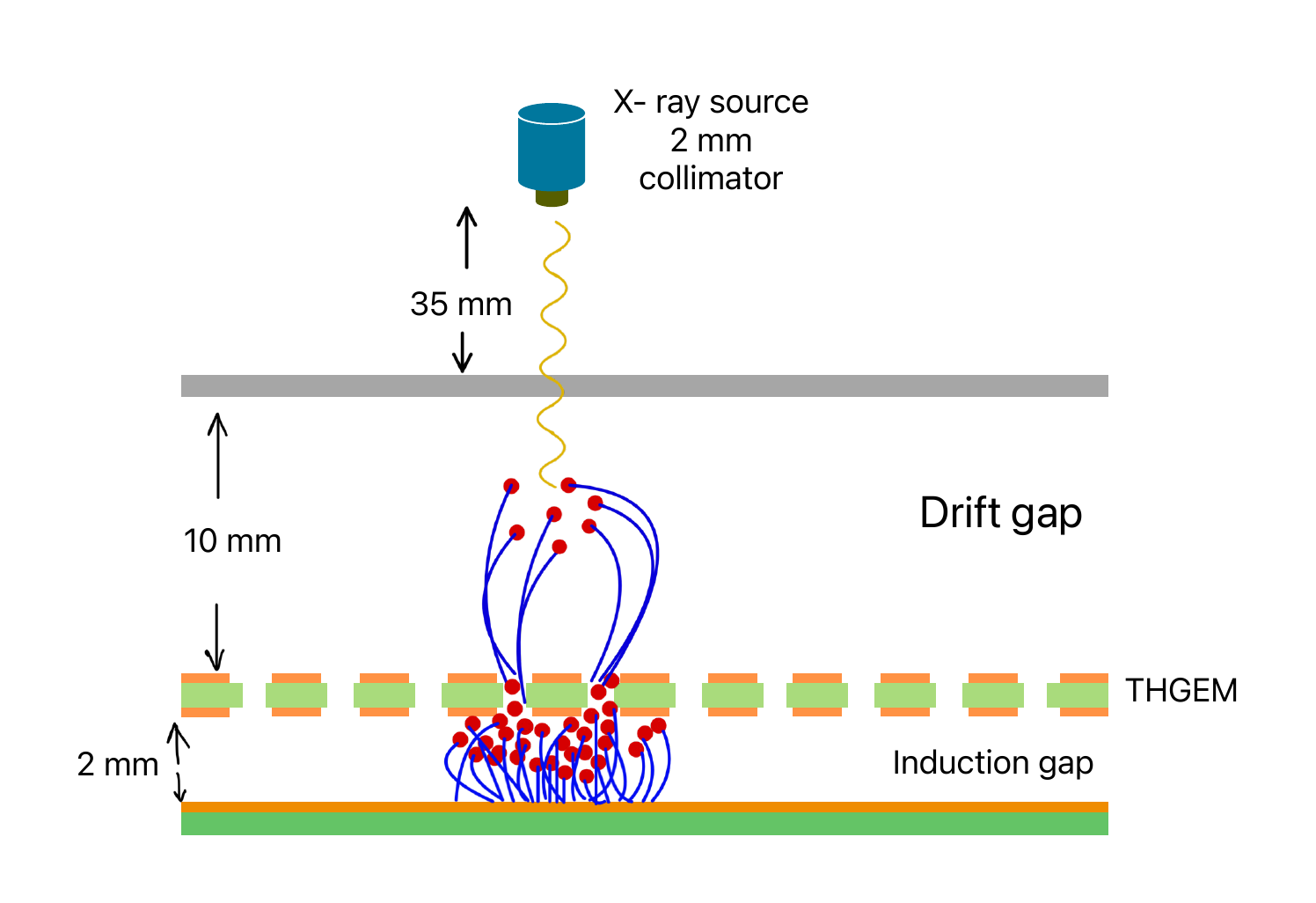}
\caption{Single-stage THGEM detector set-up\label{fig:single_THGEM_setup}}
\end{figure}

The THGEM prototype detector under investigation was set in an Aluminium-made test box with thick walls to reduce electrical noise. A drift plane made of fine Steel mesh was fixed above the foil at a distance $1\,\mathrm{cm}$ (drift gap) while the readout plane, made with a Copper plate, was placed at a distance of $2\,\mathrm{mm}$ (induction gap) below the foil. The Fe$^{55}$-source emitting $5.9\,\mathrm{keV}$ X-rays, was used for the measurement. It was collimated with a hole of $2\,\mathrm{mm}$ diameter, and placed on a mylar window of the test box, approximately at a distance $35\,\mathrm{mm}$ above the drift plane. Upon entering the gaseous volume, an X-ray photon produced primary electrons in the drift gap through ionization of gaseous molecules, and the electrons drifted towards the THGEM foil under the action of the applied field. Eventually, the electrons underwent multiplication within the holes of the foil through further secondary ionization processes caused by the intense electric field existing across the foil. The field in the induction gap then pulled the electronic avalanche towards the anode which gave rise to a signal. A schematic representation of the single-stage set-up along with the internal processes of electron multiplication is illustrated in figure~\ref{fig:single_THGEM_setup}.

\begin{figure}[htbp]
\centering
\includegraphics[width=1.0\textwidth]{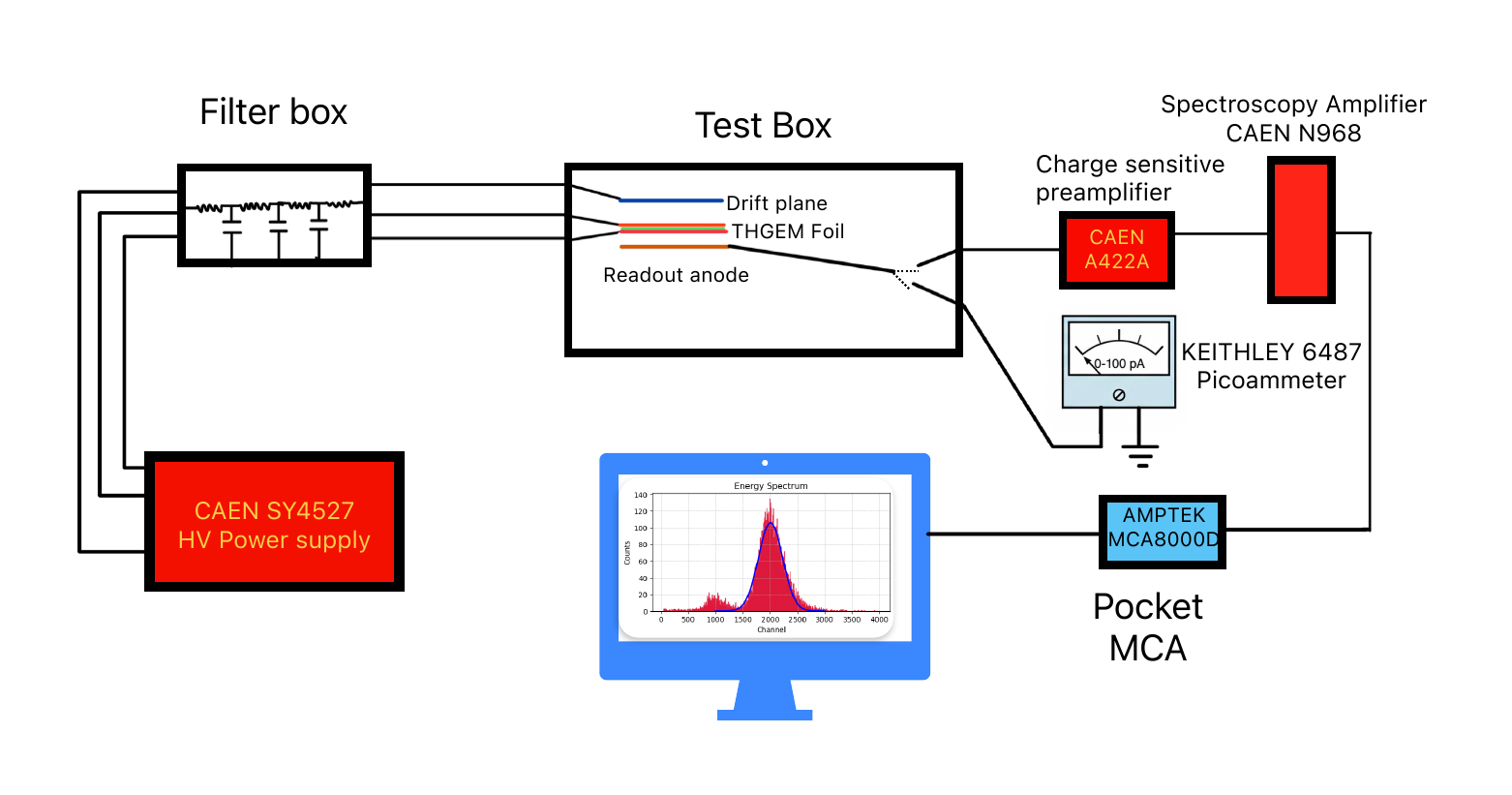}
\caption{Experimental set-up for THEGM prototype detector characterization\label{fig:test_setup}}
\end{figure}

The test box containing the THGEM prototype detector was initially purged using UHP nitrogen for approximately $24\,\mathrm{hrs}$ at a flow rate of about $7\,\mathrm{sccm}$, which turned over the internal volume at least five times. Subsequently, a high-voltage test of the THGEM foil followed by gradually ramping up the voltage supply at a rate of $5\,\mathrm{V/s}$ initially, and reducing it to a slower rate of $2\,\mathrm{V/s}$ later, and finally $1\,\mathrm{V/s}$ near the Paschen limit. In case of a spark, the detector was allowed to rest at that voltage level for at least half an hour till no further spark was observed. 
Later, the filling gas Ar-CO$_2$ (90:10) at atmospheric pressure was circulated at the same rate and for same duration of the UHP nitrogen to prepare the prototype detector for characterization tests. To operate the detector, suitable high voltages were applied to the drift plane and across the THGEM foil from CAEN SY4527 universal multichannel power supply system through a low-pass filter box. The readout anode plane was kept grounded through Keithley 6487 picoammeter and current signal was measured while ramping up the voltage to check if any high leakage current arises or discharges happen before crossing the working voltage region. The current pulse induced on the anode due to the propagation of electronic avalanche was converted to voltage by CAEN A422A charge-sensitive preamplifier and amplified in CAEN N968 spectroscopy amplifier before its acquisition by Amptek MCA8000D multichannel analyser (MCA). The diagram of the experimental set-up for the THGEM detector characterization is shown in figure~\ref{fig:test_setup}. 

\subsection{Gain characterization}
\label{Gain_char}
\begin{figure}[htbp]
\centering
    \subfloat[ \label{fig:heed_primaries}]{%
     \includegraphics[width=0.55\textwidth]{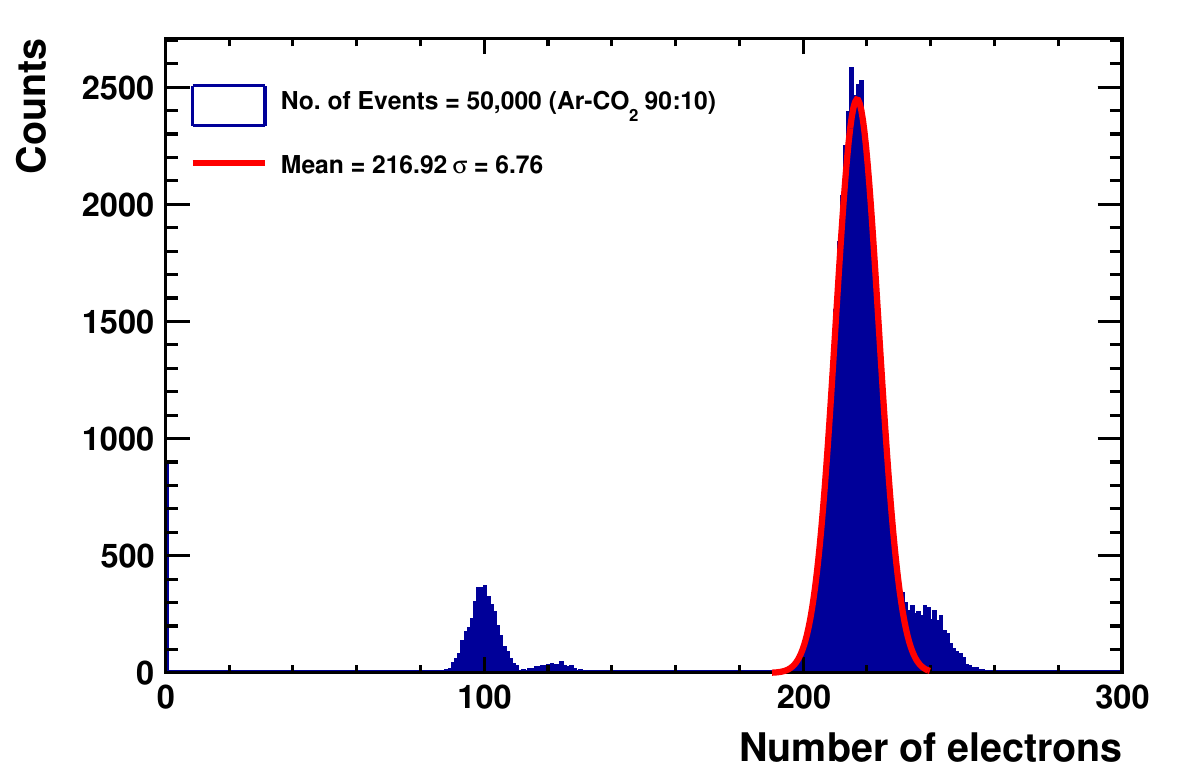}}
    \hfill
     \subfloat[ \label{fig:mca_spectrum}]{%
      \includegraphics[width=0.55\textwidth]{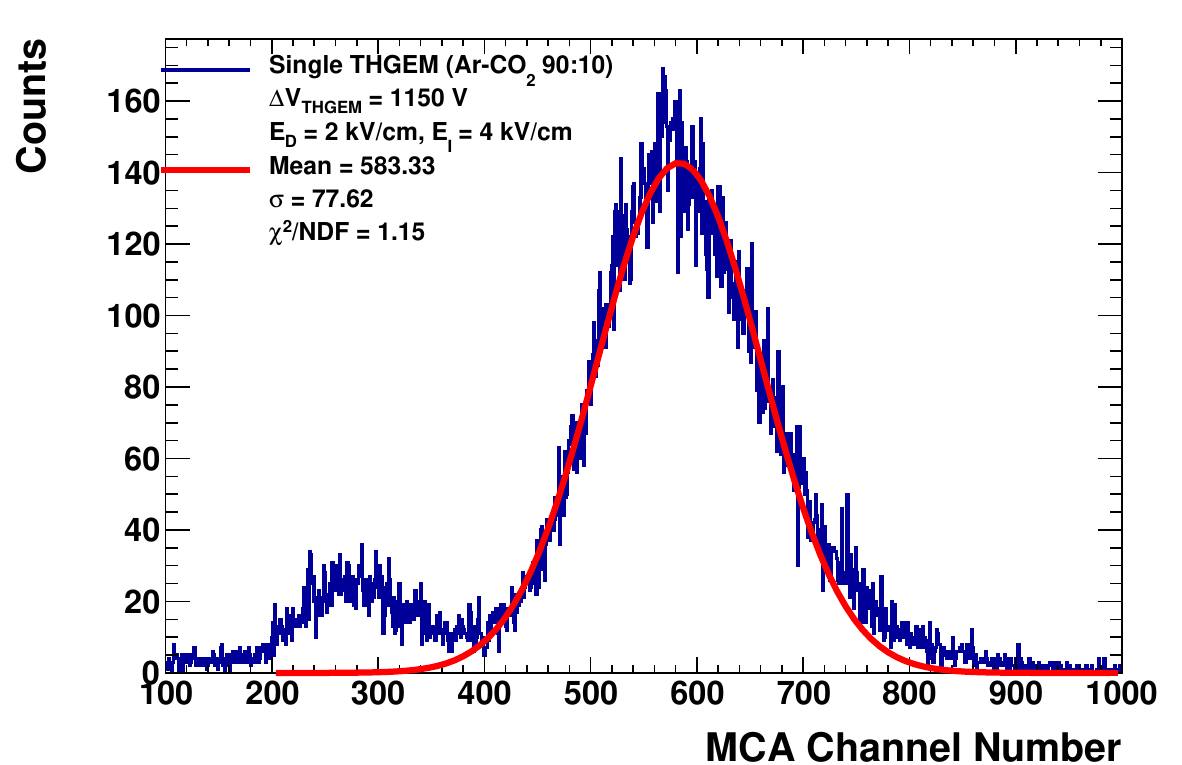}}
\caption[]{(a) Primary electrons produced by $5.9\,\mathrm{keV}$ X-ray in Ar-CO$_2$ (90:10), as obtained from \textsc{heed}~\cite{Smirnov2005}, and (b) X-ray spectrum of Fe$^{55}$-source, as acquired by the MCA}
\end{figure}

The gas gain of the THGEM detector was determined from the following expression:
\begin{equation}
\label{GG}
    G = \frac{Q_{c} / e}{N_{p}}
\end{equation}
where $Q_{c}$ and $N_p$ are the collected charge and number of primary electrons, respectively, and $e$ is the electronic charge. It defines the gain in the number of primary electrons through multiplication via secondary ionization processes. The number of the primary electrons, $N_p$, was estimated using numerical simulation while the collected charge, $Q_{c}$, was obtained from the experimental data. The $N_p$ was estimated using the \textsc{heed} toolkit~\cite{Smirnov2005} of \textsc{garfield++}~\cite{Garfield++} code. It implements \textit{Photo Absorption Ionization (PAI)} model to describe the interaction of charged particles or photons with matter by relating the cross-section of energy loss to the photoabsorption cross-section of the medium. For each X-ray event, the primary ionization clusters of localized group of electron-ion pairs, produced along the X-ray photon track in the gas medium were computed, including atomic relaxation effects like Auger electrons and fluorescence. 
The histograms of primary ionization electrons produced by 50,000 events of $5.9\,\mathrm{keV}$ X-ray in Ar-CO$_2$ (90:10) is shown in figure~\ref{fig:heed_primaries}, as an example. The mean number of the primary electrons, 217, as obtained from Gaussian fitting of the main peak, resulting from the total absorption of the $5.9\,\mathrm{keV}$ photon in the Argon K-shell, was considered as $N_p$-value for the corresponding gas mixture.

\begin{figure}[htbp]
\centering
    \includegraphics[width=0.7\textwidth]{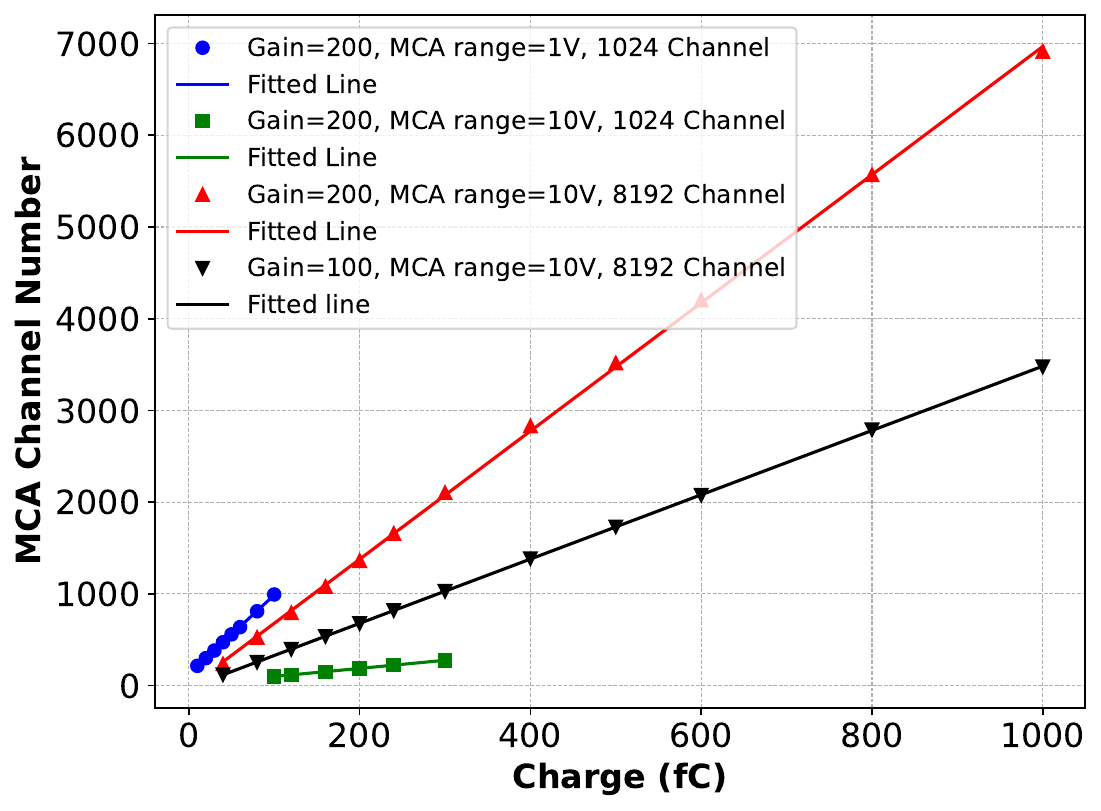} 
\caption[]{\label{fig:Calib_All} Charge calibration of MCA output}
\end{figure}

The collected charge, $Q_c$, was determined from the voltage signal, as acquired by the MCA through the readout electronics, when the THGEM detector was irradiated with $5.9\,\mathrm{keV}$ X-ray from the Fe${55}$-source.  Figure~\ref{fig:mca_spectrum} illustrates the typical X-ray pulse height spectrum, produced in the Ar-CO$_2$ (90:10), showing the main photo-peak corresponding to $5.9\,\mathrm{keV}$ along with the smaller Argon escape peak. 
The mean of the photo-peak, as obtained from the Gaussian fitting (shown by red solid line), was used to determine the collected charge, $Q_c$, for a given voltage settings of the THGEM detector by using calibration of the MCA channel to the input charge. It was done by acquiring input voltage pulses from ORTEC 419 precision pulse generator, passing through $2\,\mathrm{pF}$ capacitor with a $50\,\Omega$ resistance to ground, in the MCA. Figure \ref{fig:Calib_All} shows the calibration plots of input charge versus MCA channel number for different gains of the amplifier and MCA settings.

\begin{figure}[htbp]
\centering
     \includegraphics[width=0.7\textwidth]{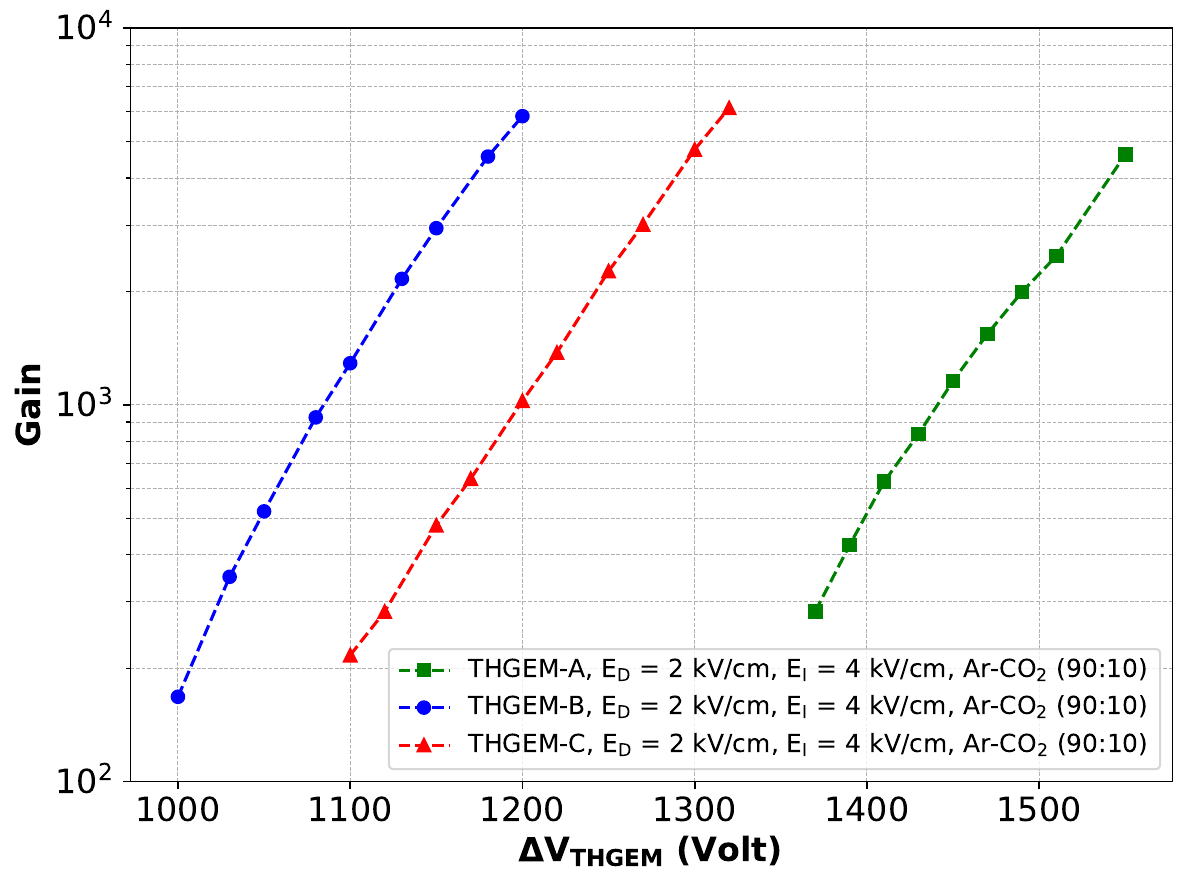}
\caption[]{\label{fig:gain_3THGEMs} Comparison of gain in Ar-CO$_2$ (90:10) of prototypes made with THGEM-A, THGEM-B, and THGEM-C}
\end{figure}

The gain characterization was carried out for the prototype detectors made with all the three THGEM foils (A, B, and C) using Ar-CO$_2$ (90:10). The result of the variation of gain with the THGEM foil voltage supply, $\Delta V_{THGEM}$, is illustrated in figure~\ref{fig:gain_3THGEMs} for the given drift field, $E_D=2\,\mathrm{kV/cm}$, and induction field, $E_I=4\,\mathrm{kV/cm}$. 
As the first Townsend coefficient, $\alpha$, depends on the electric field inside the THGEM foil holes, and the total ionization varies exponentially with the $\alpha$, an approximately linear trend was observed in the logarithmic plot of gain versus $\Delta V_{THGEM}$. This behaviour was influenced by charging-up of the THGEM foil, which eventually led to a stooping trend in the gain at higher voltages. 

\begin{figure}[htbp]
\centering
   \subfloat[ \label{fig:Single:gain}]{%
    \includegraphics[width=0.5\textwidth]{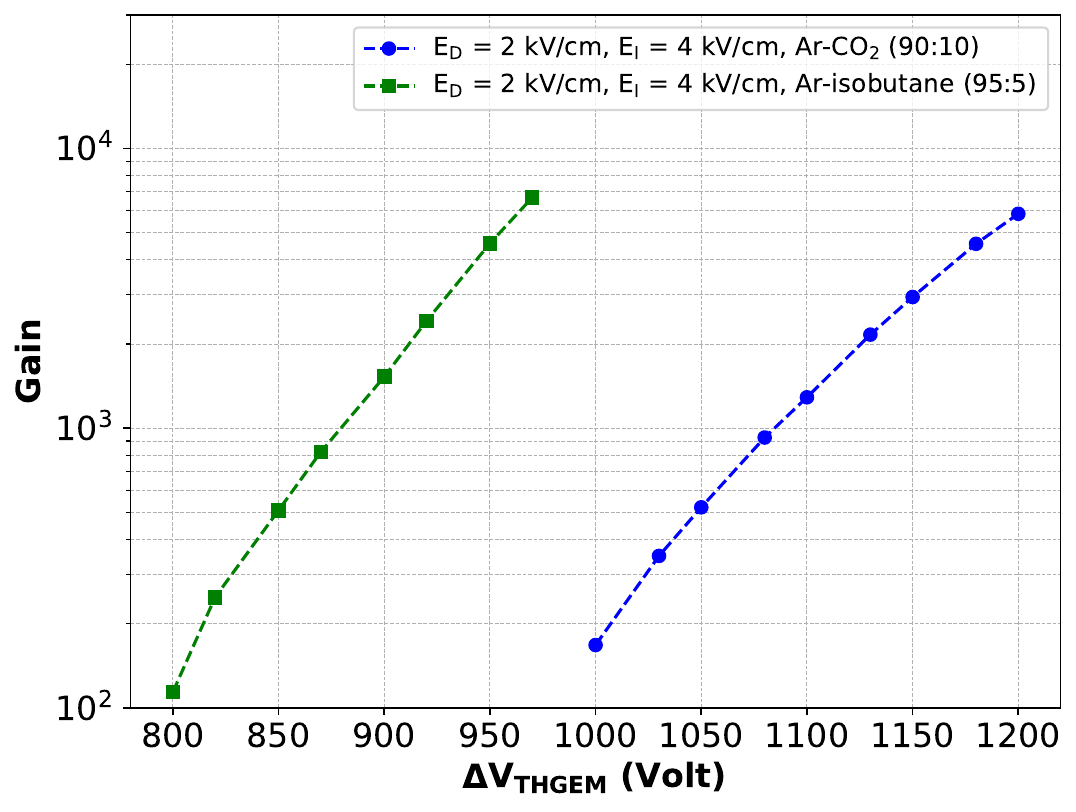}}
  \hfill
   \subfloat[ \label{fig:Single:EnRes}]{%
    \includegraphics[width=0.5\textwidth]{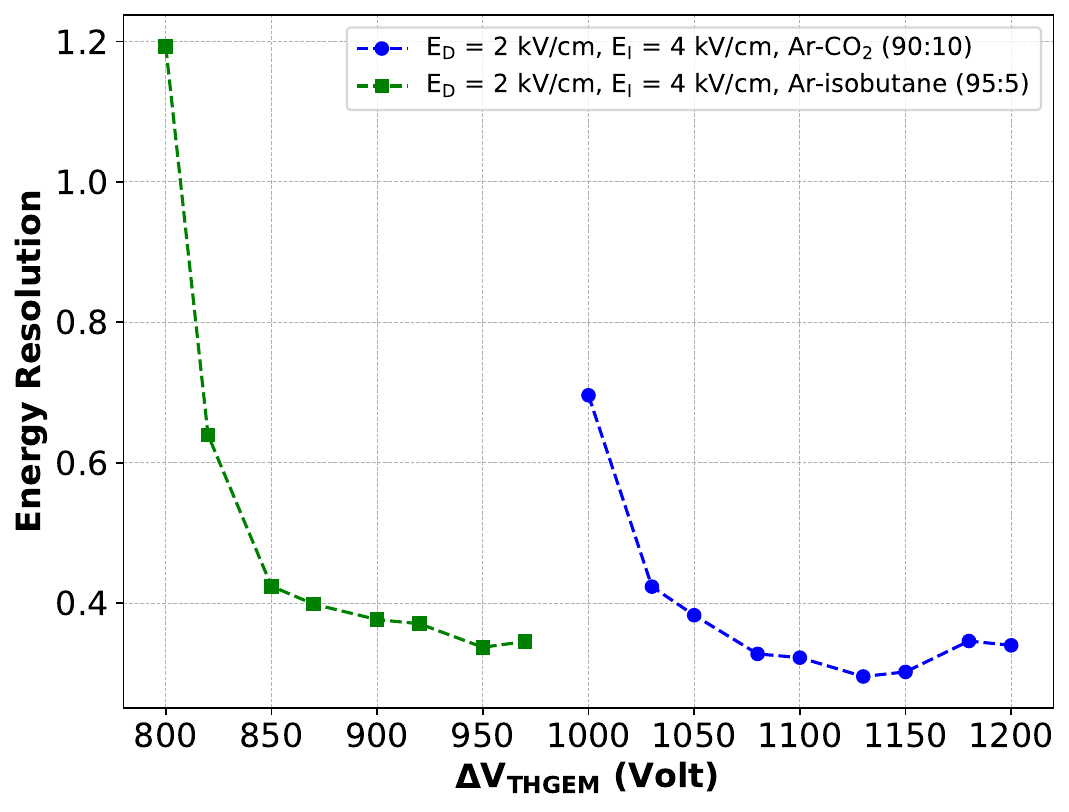}}
\caption[]{\label{fig:Single:Gain_EnRes} Measurement of (a) gain, and (b) energy resolution of single THGEM-B in Ar-CO$_{2}$ (90:10) and Ar-isobutane (95:5) at different working voltages \label{fig:gain_resolution} }
\end{figure}

On the basis of the gain study of the three THGEM prototypes, the THGEM-B prototype was considered as the optimal choice owing to its lower operating regime, and subjected to further extensive investigations. Eventually, it was investigated to characterize its performance using Ar-isobutane (95:5) as well since the choice of operating gas significantly impacts charge amplification, operational stability, and spatial resolution of the detector. The results of the gain and energy resolution measurements using both the gas mixtures are compared in figure~\ref{fig:Single:gain} and figure~\ref{fig:Single:EnRes}, respectively.

\begin{figure}[htbp]
\centering
   \subfloat[ \label{fig:Single:DriftVar}]{%
    \includegraphics[width=0.5\textwidth]{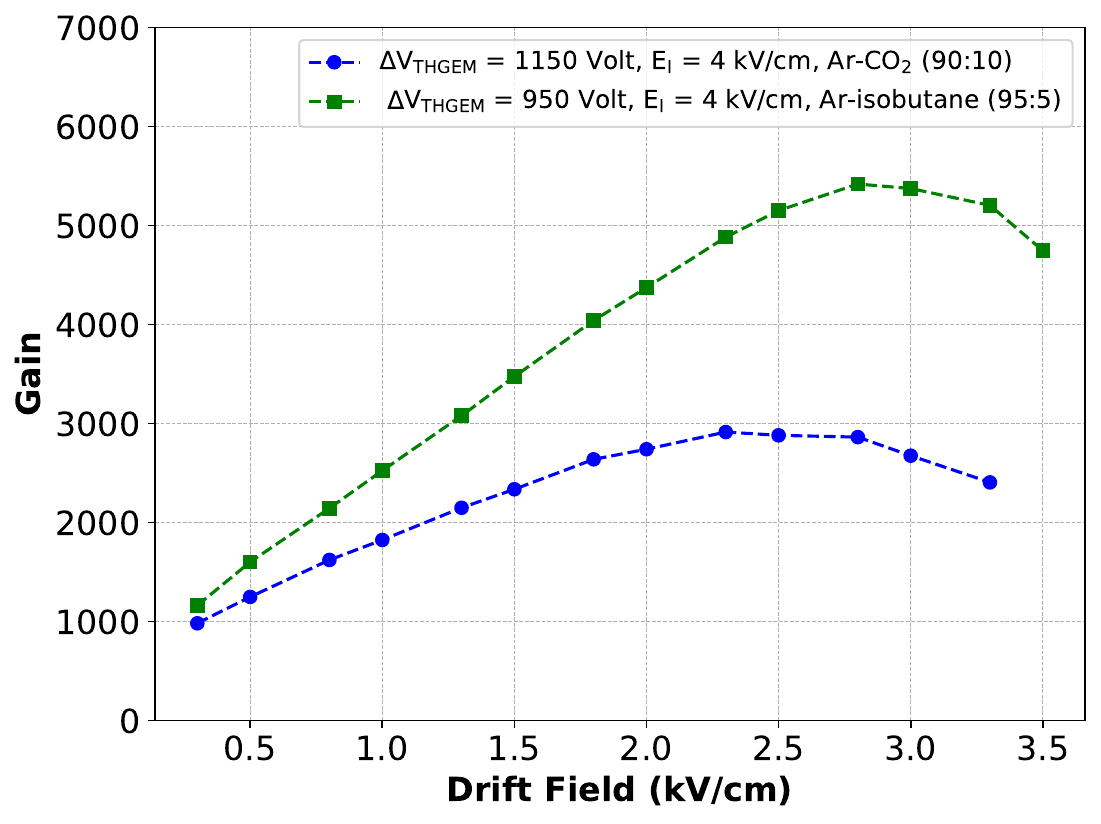}}
  \hfill
   \subfloat[ \label{fig:Single:IndVar}]{%
    \includegraphics[width=0.5\textwidth]{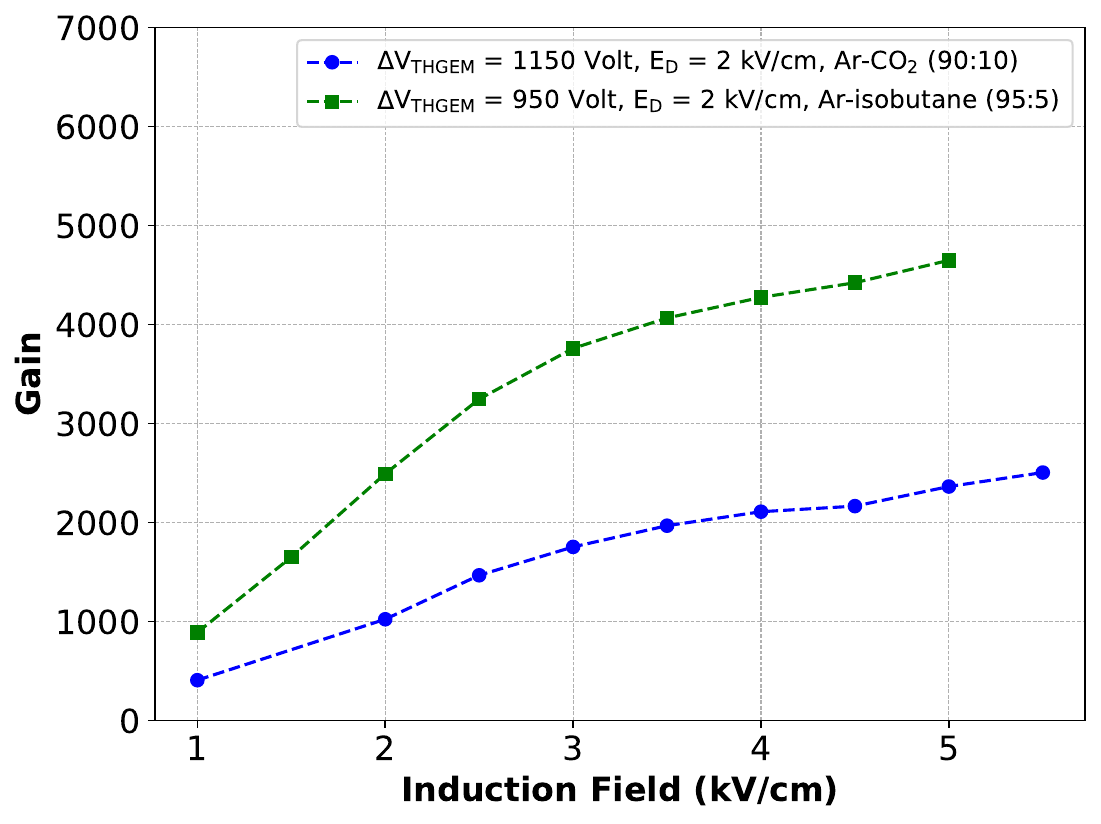}}
\caption[]{\label{fig:Single:Drift_IndVar} Gain of single-stage THGEM (THGEM-B) detector in Ar-CO$_{2}$ (90:10) and Ar-isobutane (95:5) as a function of (a) drift field, and (b) induction field}
\end{figure}

As an additional step, the effect of variation in drift and induction field values on the gas gain was investigated to optimize the detector performance. In order to find the optimal drift field value, it was varied from $0.3$ to $3.5\,\mathrm{kV/cm}$ while keeping the induction field, $E_I = 4\,\mathrm{kV/cm}$. As can be noted from figure~\ref{fig:Single:DriftVar}, the gain in both of the gas mixtures increased with the rise in drift field to reach a maximum value, and eventually, decreased at higher drift field values above $2.5\,\mathrm{kV/cm}$. The possible reason can be explained as follows. With the increase in the drift field, electron collection into the THGEM holes was initially improved, leading to a rise in the gain. However, at sufficiently high drift fields, the field configuration became less effective in focusing the electrons into the holes, and a fraction of the electrons was lost on the top surface of the THGEM foil, resulting in a drop in the gain. In the same manner, the induction field was raised from $1.0$ to $5.0\,\mathrm{kV/cm}$ when the drift field was fixed at $E_D = 2\,\mathrm{kV/cm}$. Figure~\ref{fig:Single:IndVar} shows that the gain increased rapidly at low induction fields due to more efficient extraction of avalanche electrons and reduced loss at the bottom copper electrode. This was followed by a region between $3.0$ to $4.0\,\mathrm{kV/cm}$ where the gradient of the gain curve was relatively decreased. 

By comparing the detector performance with both the Ar-based mixtures, it can be inferred that the Ar-isobutane (95:5) is a better choice as it offers a working voltage regime lower than that of Ar-CO$_2$ (90:10) facilitating safe operation with less charging-up issue, and achieves better gain.

\subsection{Double-stage operation}
\label{DTHGEM}
\begin{figure}[htbp]
\centering
\includegraphics[width=.7\textwidth]{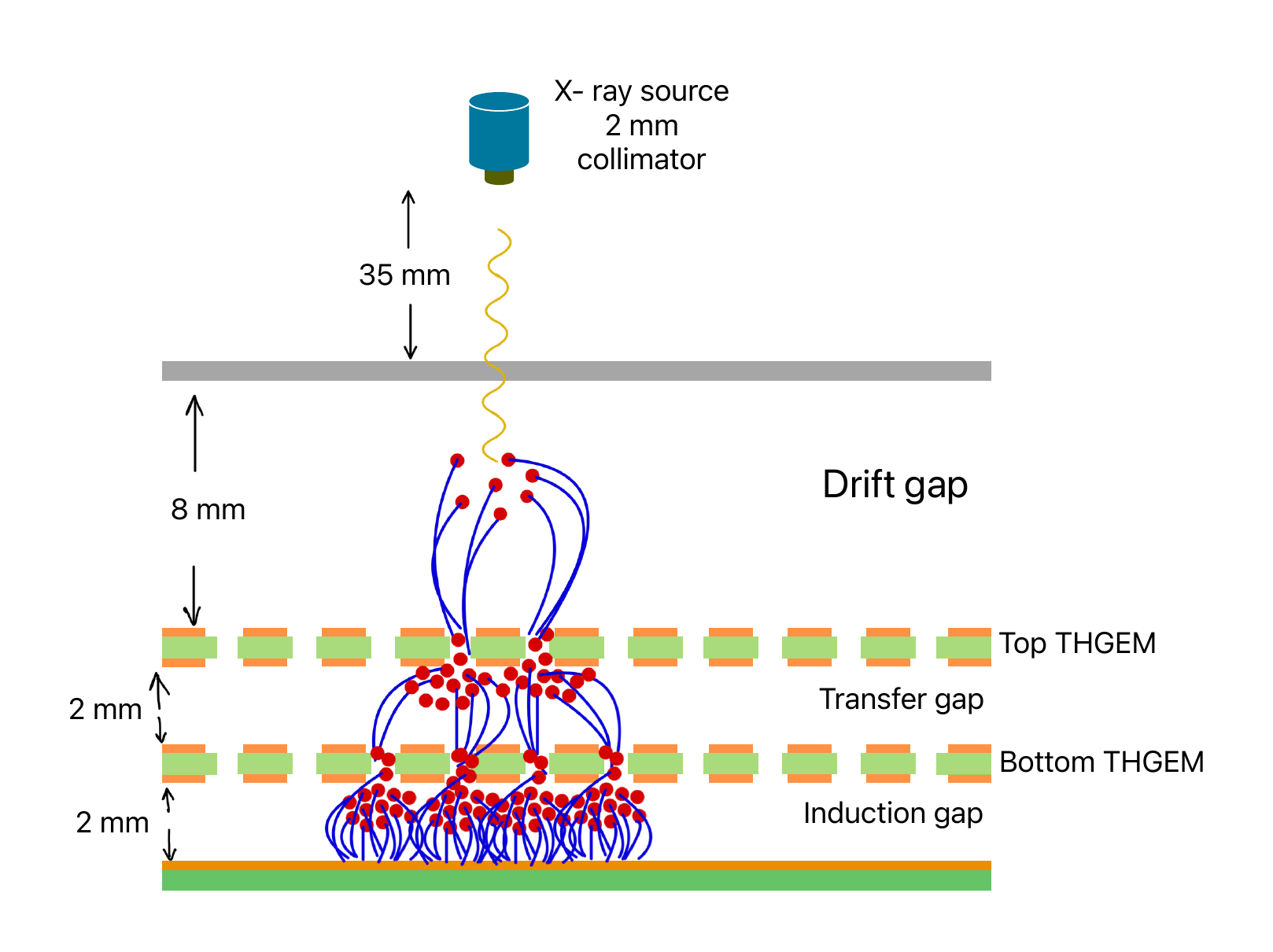}
\caption{Double-stage configuration and operation of THGEM detector\label{fig:double_layer_schematic}}
\end{figure}

To increase the gain while reducing the probability of discharge on the detector foil, the multi-stage operation of the THGEM detector was explored. A double-stage prototype was built using THGEM-B foils where an additional transfer gap was produced in-between two THGEM foils.
The drift gap, transfer gap and induction gap used in this study are $8\,\mathrm{mm}$, $2\,\mathrm{mm}$, and $2\,\mathrm{mm}$, respectively. The electronic avalanche produced by the multiplication of primary electrons within the upper foil holes went through another multiplication phase in the lower foil, resulting in a cascaded amplification process. The prototype configuration and the electronic multiplication processes are schematically illustrated in figure~\ref{fig:double_layer_schematic}.
Due to electron diffusion after the first amplification stage, the second multiplication occurred in a larger number of holes. As a result, the total final charges were distributed among nearby holes, reducing the charge density in each hole which eventually decreased the probability of discharges caused by a large avalanche in a single hole. 

\begin{figure}[htbp]
\centering
   \subfloat[ \label{fig:Doublegem_gain}]{%
    \includegraphics[width=0.5\textwidth]{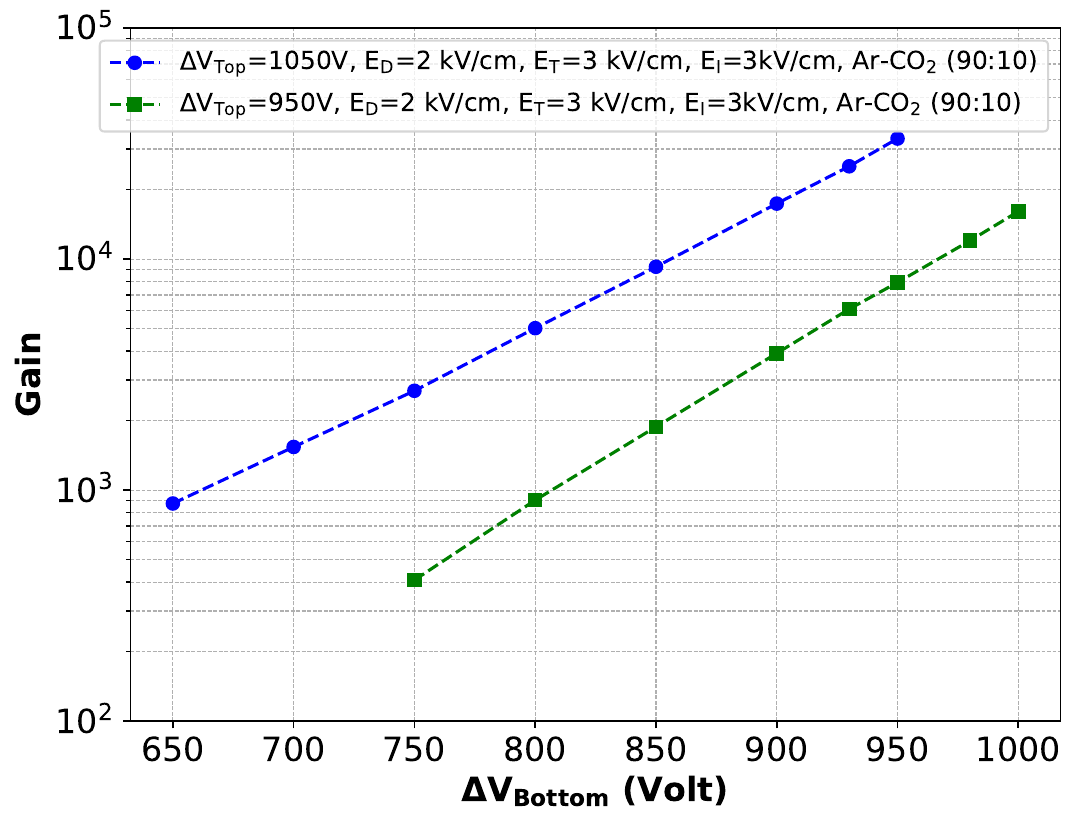}}
  \hfill
   \subfloat[ \label{fig:Doublegem_energyres}]{%
    \includegraphics[width=0.5\textwidth]{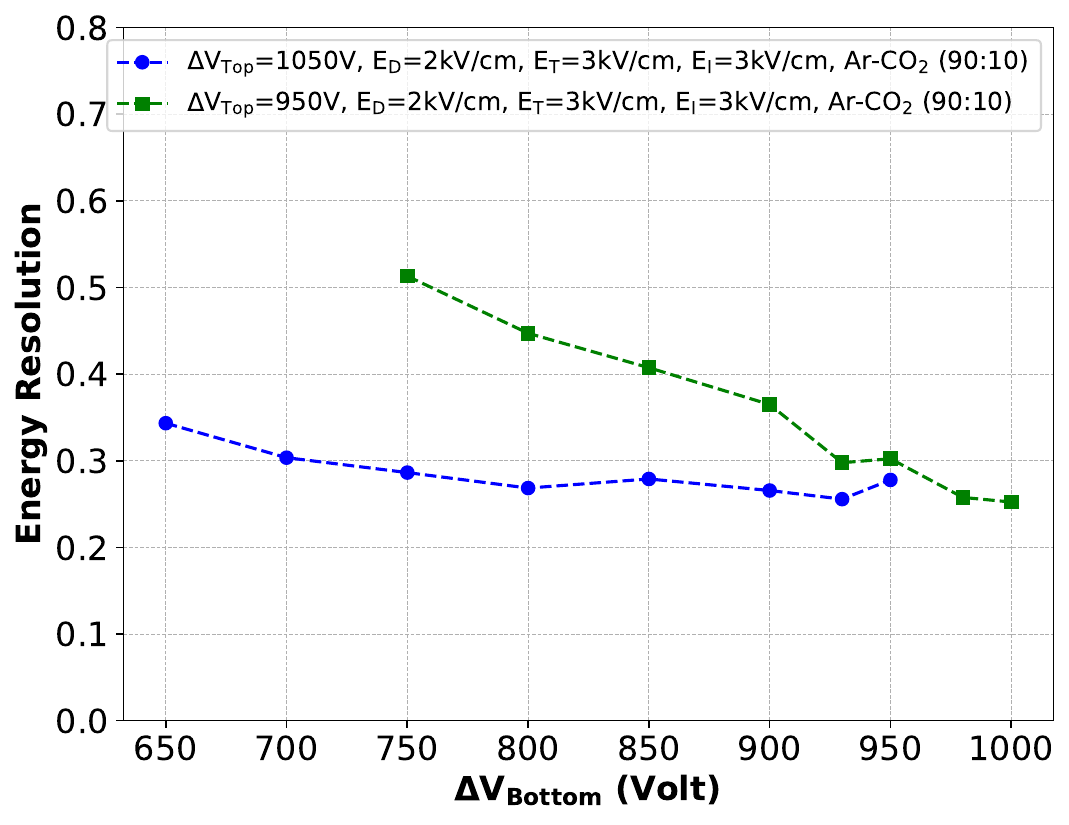}}
\caption[]{(a) Gain, and (b) energy resolution for double-stage THGEM (THGEM-B) operation using Ar-CO$_{2}$ (90:10)  }
\label{fig:Doublegem_gain_energyres} 
\end{figure}

In the present study, the voltage difference across the top THGEM, $\Delta V_{Top}$, was kept fixed at two different values $950$ and $1050\,\mathrm{V}$, while that at the bottom stage, $\Delta V_{Bottom}$, was increased gradually, and the gas gain was measured using Ar-CO$_2$ (90:10). Here, the field values at the drift, transfer and induction gaps were fixed at $E_D = 2\,\mathrm{kV/cm}$, $E_T = 3\,\mathrm{kV/cm}$, and $E_I = 3\,\mathrm{kV/cm}$, respectively. For both the cases of fixed $\Delta V_{Top}$, an exponential increase in gain was observed as the $\Delta V_{Bottom}$ was raised. A maximum value of gain around 33,000 could be achieved with $\Delta V_{Top} = 1050\,\mathrm{V}$, as can be seen in figure~\ref{fig:Doublegem_gain}.

The result of the energy resolution measurement is plotted in figure~\ref{fig:Doublegem_energyres} which was observed to improve with increase in the detector gain to a value $25-30\%$. The change is more noticeable in case of lower $\Delta V_{Top} = 950\,\mathrm{V}$. All the measurements were carried out using Ar-CO$_2$ (90:10) at atmospheric pressure.

\section{Muon detection efficiency}
\label{Efficiency} 
The muon detection efficiency of the THGEM (THGEM-B) detector was determined in reference to that of the plastic scintillators and can be defined as follows:
\begin{equation}
\label{eqn:Eff}
    \epsilon = \frac{NC_{Th}}{NC_{Sc}} \times 100\%
\end{equation}
where $NC_{Th}$ and $NC_{Sc}$ are numbers of muon events counted by the THGEM prototype detector and the scintillators, respectively. The count, $NC_{Sc}$ was obtained with the coincidence-mode operation of several scintillators.

\subsection{Experimental set-up}
\label{exp_setup}
To accomplish the efficiency measurement, a muon telescope was built using three plastic scintillators which were operated in coincidence mode to provide a trigger for recording muon events from the THGEM detector within a specific time window. The events confirmed by four-fold coincidence of three scintillators and the THGEM prototype detector together were recorded as true muon event. 

\begin{figure}[htbp]
\centering
   \subfloat[\label{fig:eff_setup_schematic}]{%
    \includegraphics[width=0.4\textwidth]{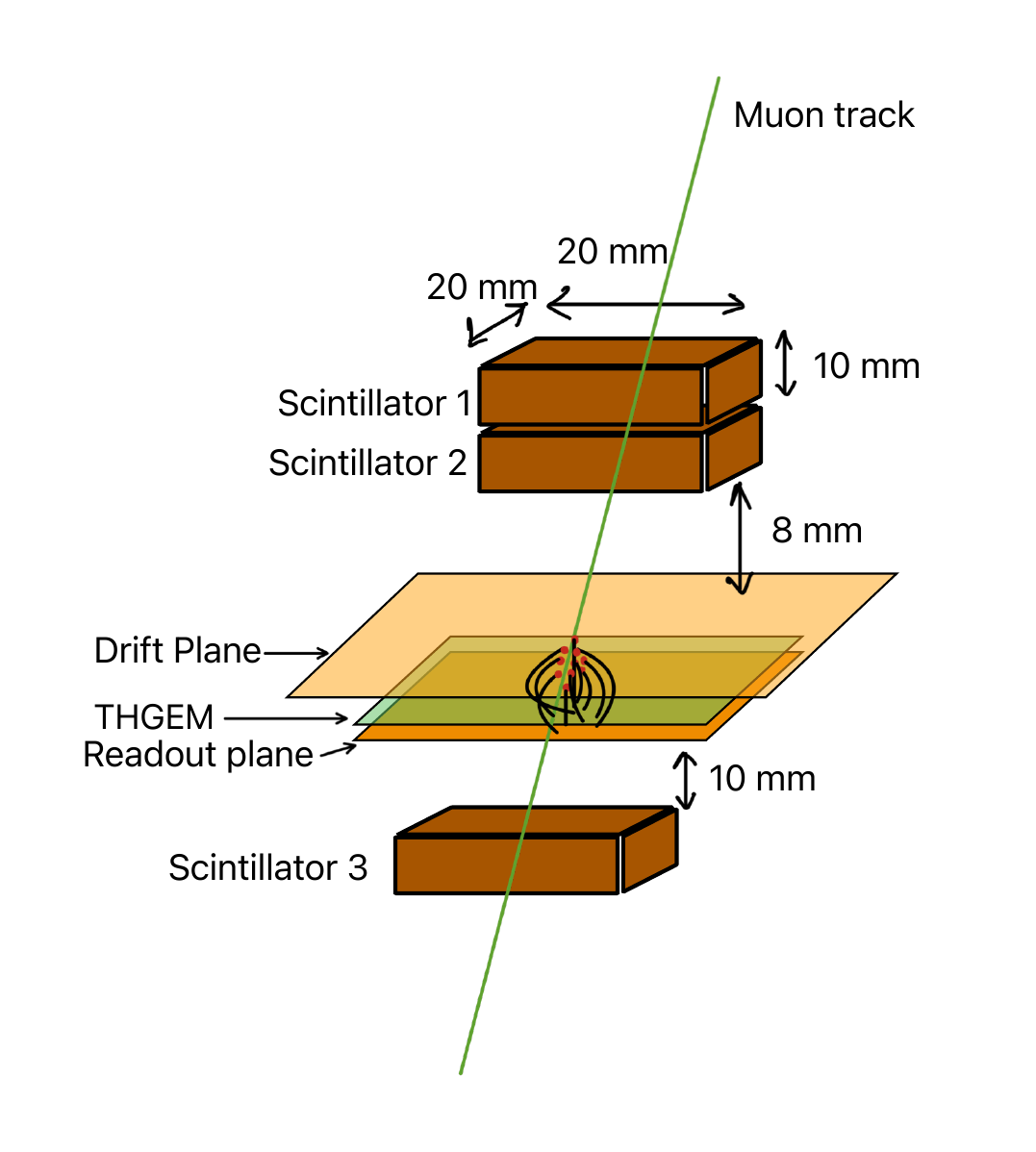}}
  \hfill
   \subfloat[ \label{fig:eff_setup_expt}]{%
    \includegraphics[width=0.6\textwidth]{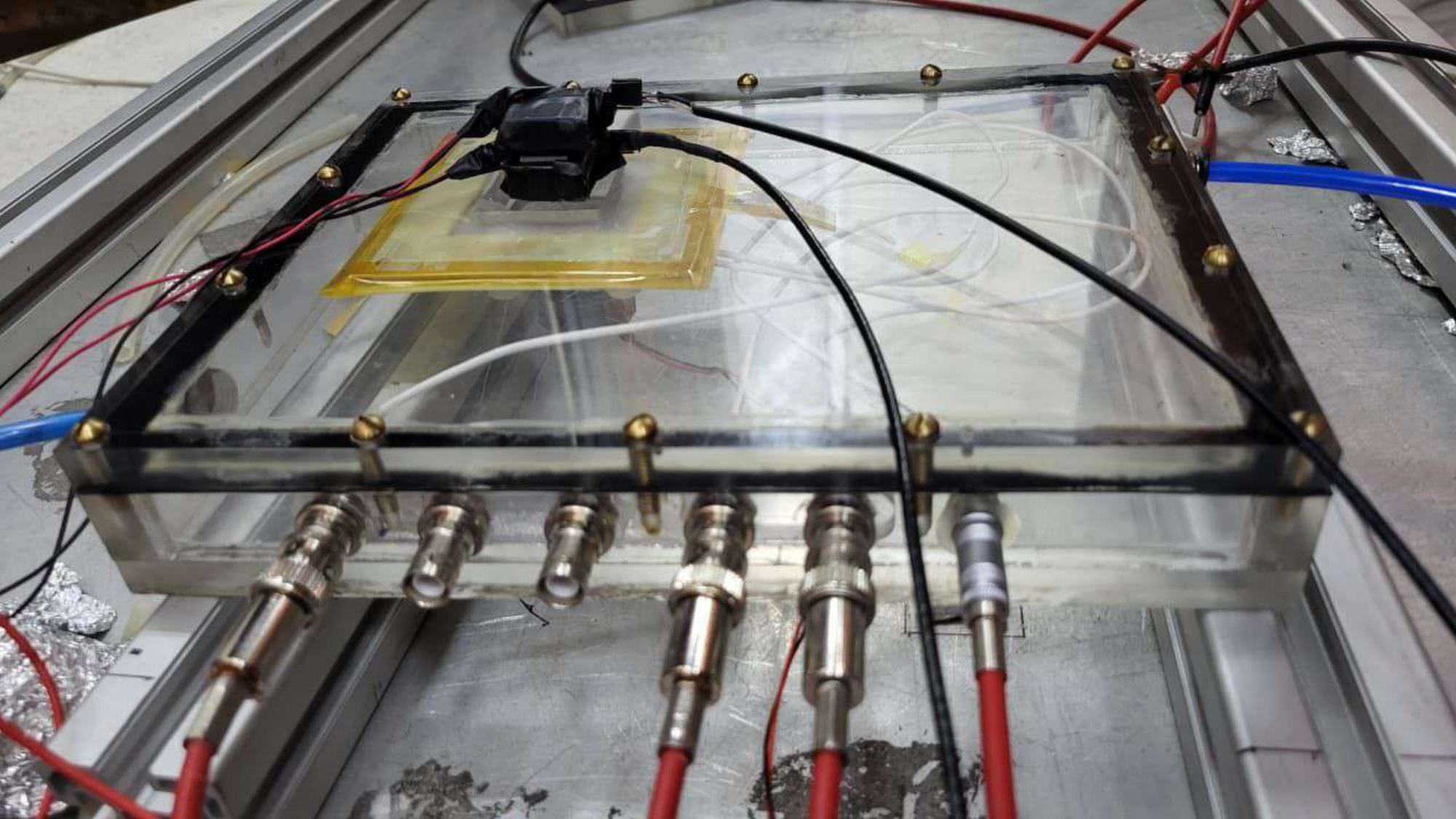}}
\caption[]{(a) Muon efficiency measurement set-up, and (b) its image}
\label{fig:readout} 
\end{figure}

To build this setup, small-area plastic scintillators of dimension $2\,\mathrm{cm} \times 2\,\mathrm{cm}$ were fabricated and coupled to SiPMs (MICROFC-60035-SMT) with active sensor area $6\,\mathrm{mm} \times 6\,\mathrm{mm}$. To maximize the angular acceptance which requires optimized distances between the scintillators and the detector planes, a Perspex-made test chamber with a height of $42\,\mathrm{mm}$ was used to house the THGEM detector. In case of single-stage operation, the drift gap was maintained at $9\,\mathrm{mm}$ and it was $6.8\,\mathrm{mm}$, in double-stage. A couple of scintillators was placed $8\,\mathrm{mm}$ above the drift plane, and a single one $10\,\mathrm{mm}$ below the anode readout, all aligned to overlap with the active area of the THGEM foil. The experimental setup is illustrated schematically in figure~\ref{fig:eff_setup_schematic} along with its image shown in figure~\ref{fig:eff_setup_expt}. 

The SiPMs were powered using an Agilent E3634A low-voltage power supply by applying a bias voltage of $29\,\mathrm{V}$ through a suitable RC filter, corresponding to approximately $5\,\mathrm{V}$ above the breakdown voltage. The signals from three SiPMs were collected in a Tektronix MDO34 oscilloscope with a set threshold of $5\,\mathrm{mV}$ and configured with AND logic to generate a trigger of possible muon event. The signal from the THGEM detector readout was collected in the oscilloscope following its processing through CAEN A422A charge sensitive preamplifier and N968 spectroscopy amplifier, as shown in figure~\ref{fig:test_setup}. Here, a threshold of $50\,\mathrm{mV}$ was set to restrict the noise. 

\subsection{Experimental procedure}
The THGEM prototype was operated with different voltage settings using the gas mixture Ar-CO$_2$ (90:10). The detector signal was delayed by about $1.65\,\mathrm{\mu s}$ relative to the scintillator pulses due to signal pre-processing. When all three scintillators produced positive signals crossing the threshold within a time window of $8\,\mathrm{ns}$, a trigger of three-fold coincidence of the scintillators with a window width $20\,\mathrm{\mu s}$ was generated. The THGEM signals, reaching within the trigger window, were recorded to store a true muon event. 
The measurement was carried out for both the single and double-stage configurations. For each voltage configuration, data were collected for $14-16\,\mathrm{hrs}$, yielding about 800 total recorded events. 

\subsection{Analysis and results}
\label{analysis}
\begin{figure}[htbp]
\centering
\includegraphics[width=.7\textwidth]{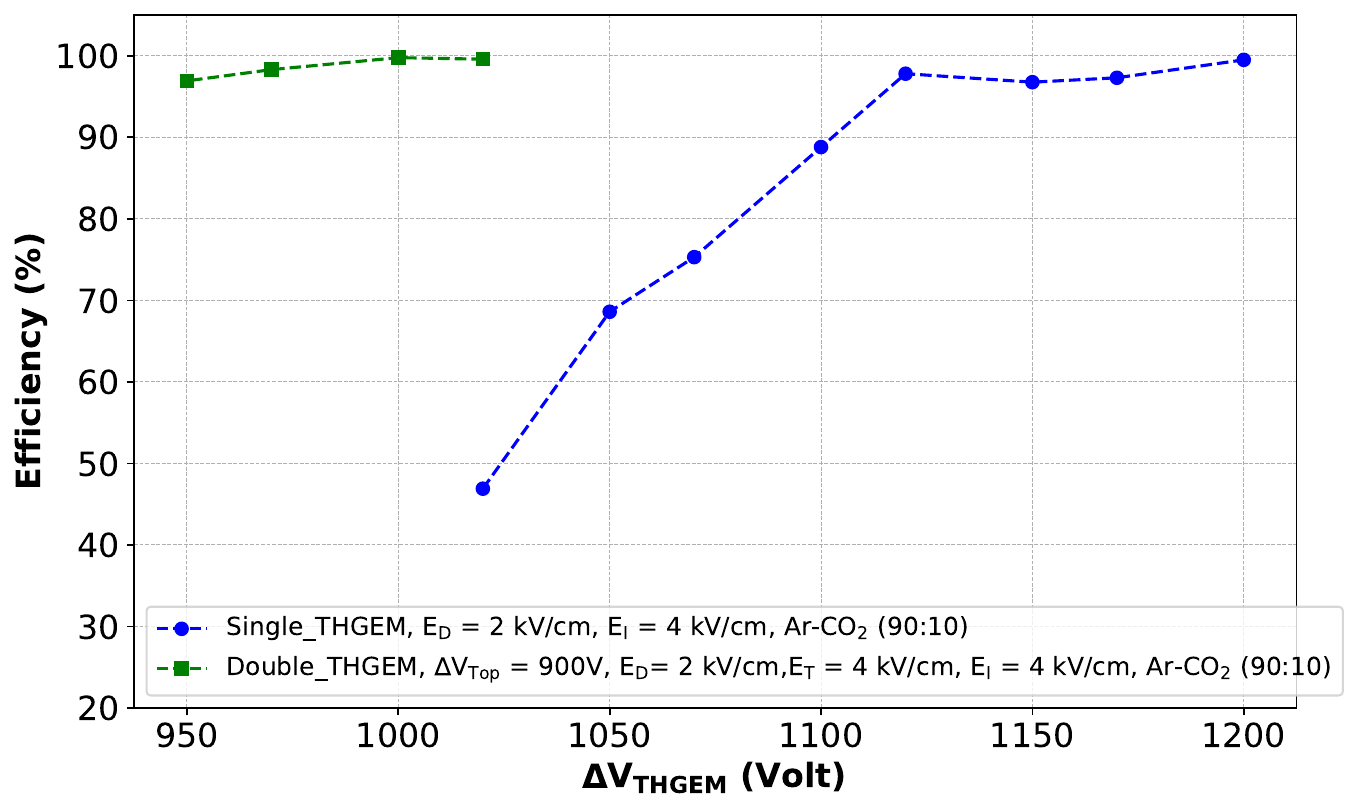}
\caption{Muon detection efficiency of single and double-stage THGEM detectors\label{fig:muon_efficiencyplot}}
\end{figure}

Some of the recorded events were found to contain noise signals generated within the scintillators, or electrical noise induced in the scintillators due to sparks in the THGEM detector. However, these raw noise pulses from the scintillators were filtered after investigating their pulse shapes which were different from those of true signal pulses. After rejecting noise and spark-induced events, around 600–700 events were used to calculate the muon efficiency. The muon detection efficiency of the THGEM detector was calculated following the equation~\ref{eqn:Eff} where the number of muon events recorded by it, $NC_{Th}$, was determined by analysing a $1\,\mathrm{\mu s}$ wide window after $1.15\,\mathrm{\mu s}$ within the trigger window. The total number of muon events detected by the scintillators, $NC_{Sc}$, was obtained from the total number of trigger events, recorded in the data-set, which is about 600-700, as mentioned before.
Figure~\ref{fig:muon_efficiencyplot} illustrates the results of the muon efficiency measurement with single and double-stage operation of the THGEM prototype detectors using Ar-CO$_2$ (90:10) as a function of $\Delta V_{THGEM}$. In case of the double-stage configuration, the $\Delta V_{Top}$ was kept fixed while $\Delta V_{Bottom}$ was varied (represented by $\Delta V_{THGEM}$, here). In both the cases, muon detection efficiency was found to be more than 95\% over a substantially long range of THGEM supply voltage, $\Delta V_{THGEM}$.
 
\section{Position resolution}
\label{Resolution}
The position resolution of the THGEM prototype detector, equipped with a multi-strip readout geometry, was measured following a non-traditional technique, as reported by V. Kumar \textit{et al}~\cite{VKumar2024}. It utilized a high-precision 3-axis positioning device to place a collimated Fe$^{55}$-source to irradiate the THGEM detector and the resolution was determined from the standard deviation of the error in detecting the position of the event caused by $5.9\,\mathrm{keV}$ X-ray, emitted from the Fe$^{55}$-source, by the THGEM detector while the actual location of the event was provided by the 3-axis positioning device from the placement of the source.

\subsection{Readout design}
\begin{figure}[htbp]
\centering
   \subfloat[ \label{fig:readout_img}]{%
    \includegraphics[width=0.5\textwidth]{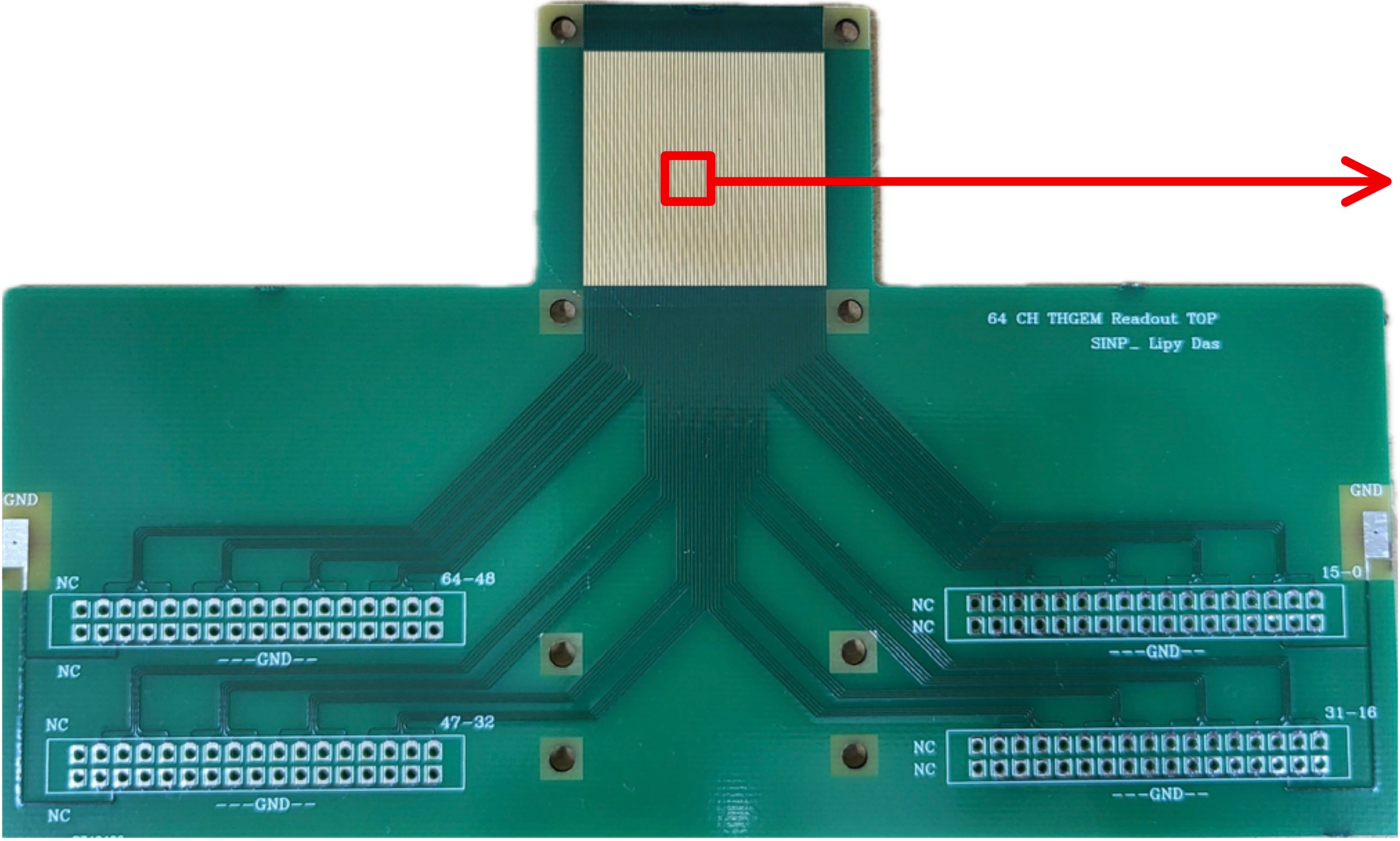}}
  \hfill
   \subfloat[ \label{fig:microsope_readout}]{%
    \includegraphics[width=0.5\textwidth]{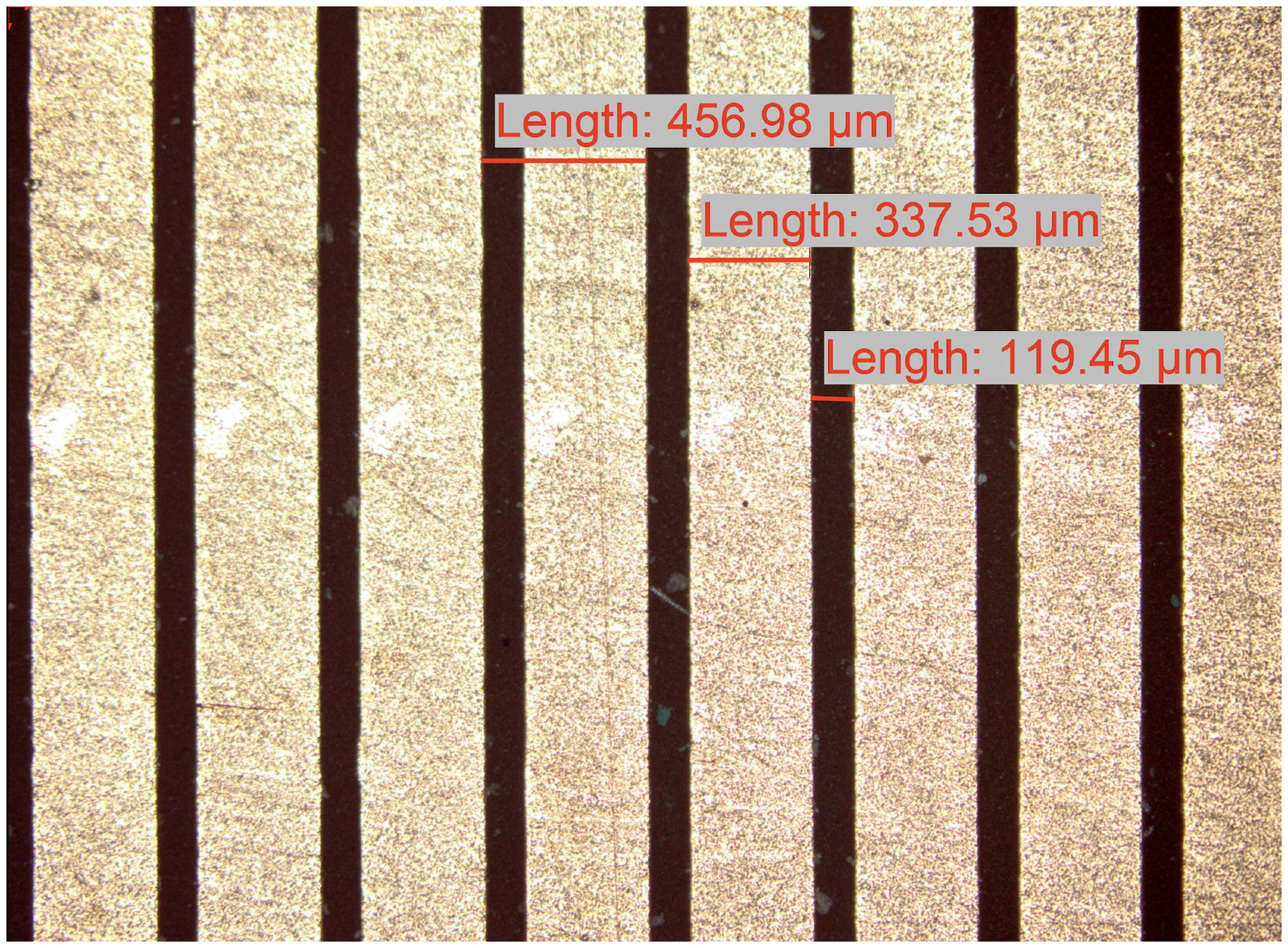}}
\caption[]{(a) Readout PCB, and (b) the readout strips (5X magnification)}
\label{fig:readout} 
\end{figure}

The previously used continuous readout plane in the THGEM prototype was replaced with a multi-strip anode PCB, designed at the laboratory and fabricated from a local PCB manufacturing company. The same is depicted in figure~\ref{fig:readout_img}. It is a double-layer PCB with 64 parallel readout strips, as shown in figure~\ref{fig:microsope_readout}. The strip width is $337\,\mathrm{\mu m}$ and the pitch is $457\,\mathrm{\mu m}$, as obtained from the measurement using a calibrated optical microscope with 5X magnification.
In single-stage configuration, the readout PCB was placed at $2\,\mathrm{mm}$ below the THGEM foil while the drift gap was set to $8\,\mathrm{mm}$. For double-stage operation, the drift, transfer, and induction gaps were set to $6\,\mathrm{mm}$, $2\,\mathrm{mm}$, and $2\,\mathrm{mm}$, respectively. 

\subsection{Experimental set-up}
\label{exp_setup_pos}
\begin{figure}[htbp]
\centering
\includegraphics[width=0.4\textwidth]{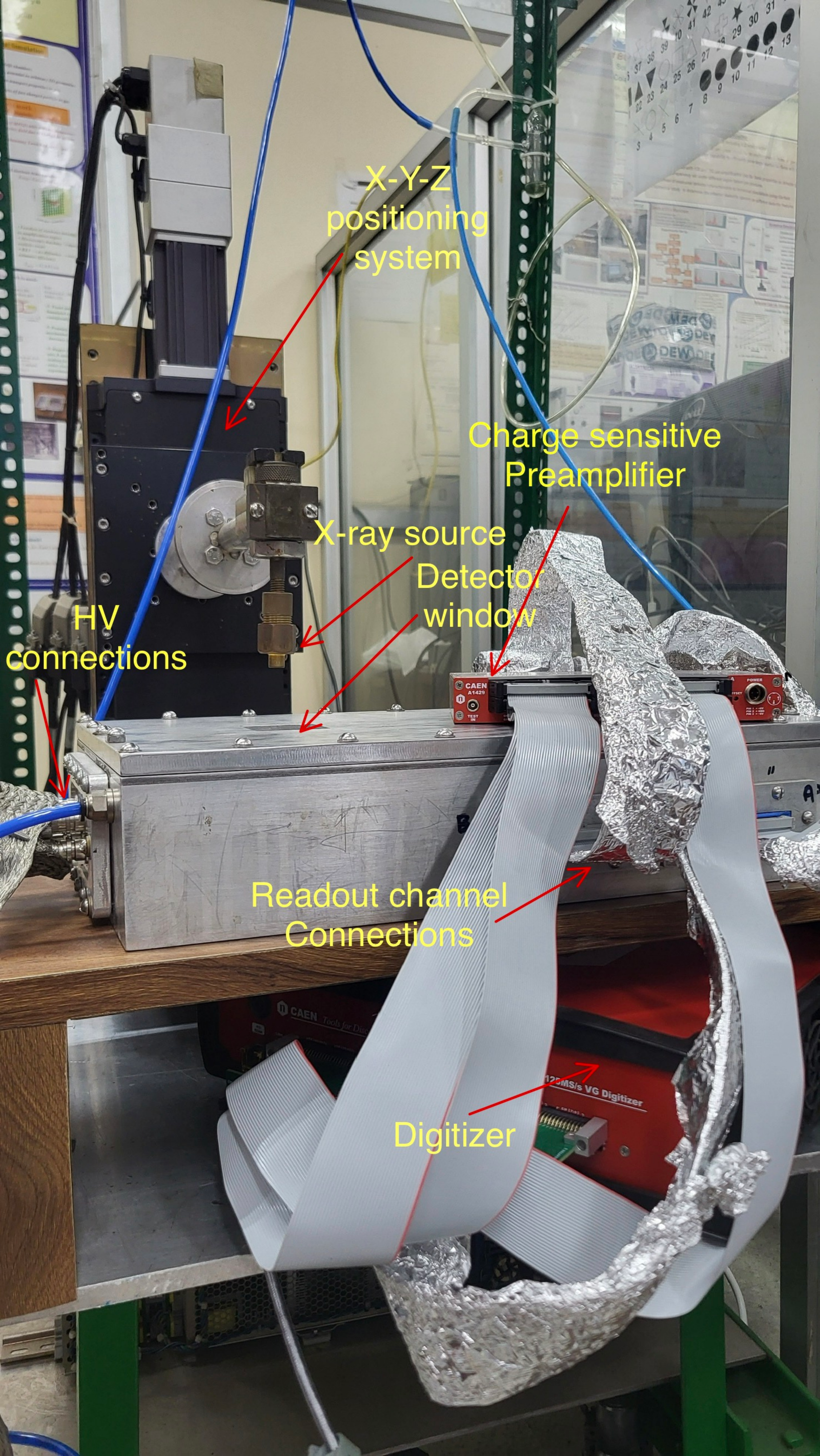}
\caption{Experimental set-up for position resolution measurement\label{fig:posres_CNC}}
\end{figure}

A high-precision 3-axis positioning system, consisting of AEROTECH ATS 165 and AEROTECH PRO 165 X-Y-Z motion stages, was utilized in the experimental set-up. It provides a travel range of $26.55\,\mathrm{cm}$ in both the X and Y-directions in the horizontal plane, while $57\,\mathrm{mm}$ vertically in the Z-direction with positioning accuracy $0.5\,\mathrm{\mu m}$. The stages were operated with AeroBasic programs, built on Aerotech Ensemble Motion Composer software platform. 
The THGEM prototype with the multi-strip readout was assembled following the single and double-stage configurations in an Aluminium-made test box with a $50\,\mathrm{\mu m}$ thick mylar window on the top. The readout plane was connected to a 64-channel, 16-bit, $125\,\mathrm{MS/s}$ ADC CAEN digitizer DT2745 through A1429 CAEN charge sensitive preamplifier. 
The DAQ system was operated using CAEN COMPASS software, through which data acquisition, online histogram plotting, and sequential data-taking runs were performed. The collimated Fe$^{55}$-source 
was mounted on the Z-axis positioning arm approximately $6\,\mathrm{mm}$ above the drift plane. Figure~\ref{fig:posres_CNC} shows the experimental arrangement containing the X-ray source mounted on the Z-axis arm of the positioning device, THGEM prototype detector in the Aluminum test box, and associated electronics.

\subsection{Experimental procedure}
\begin{figure}[htbp]
\centering
\includegraphics[width=0.6\textwidth]{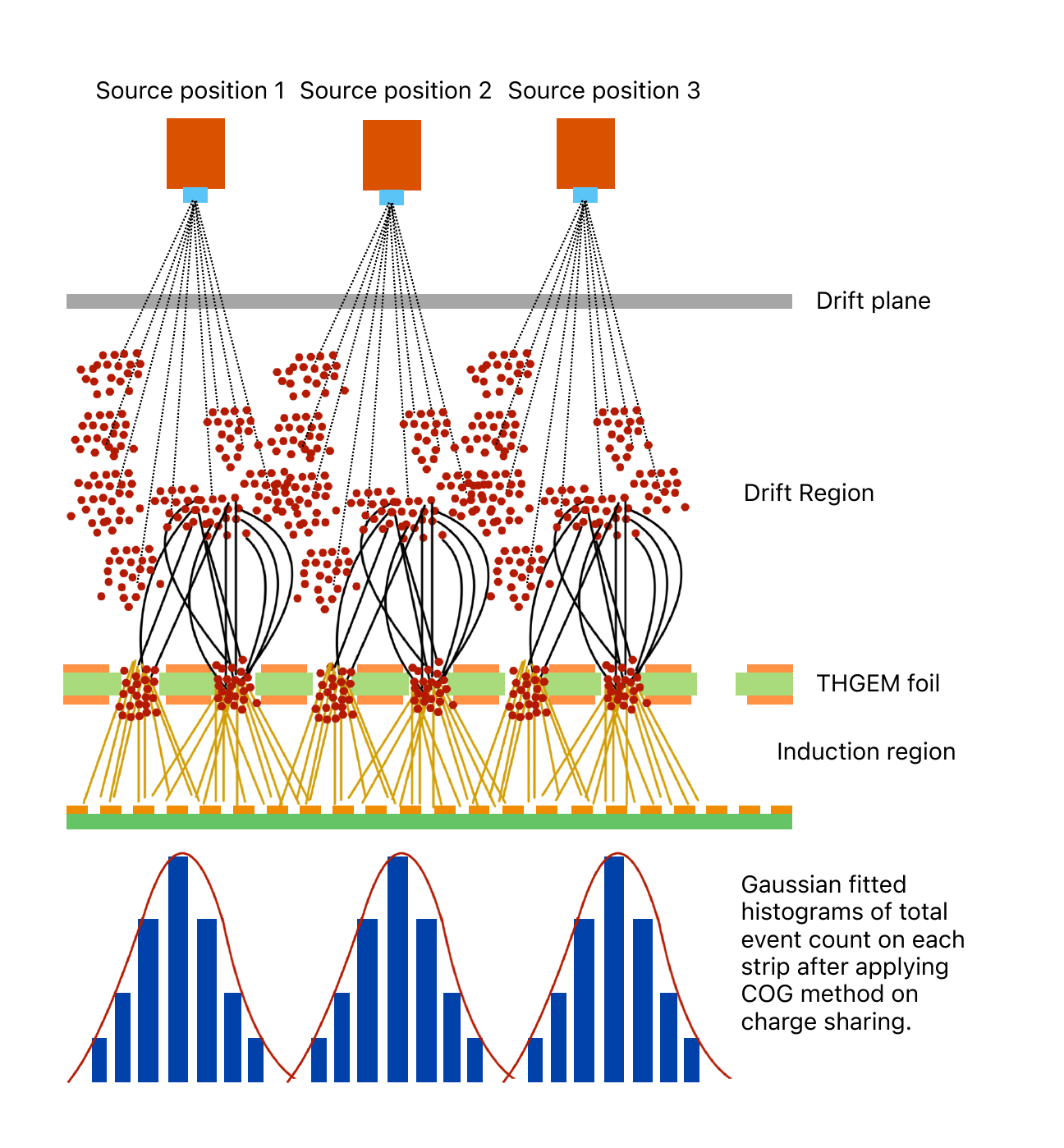}
\caption{Experimental procedure used for measurement of position resolution\label{fig:sourcemovement}}
\end{figure}

The measurement of position resolution was performed for both single and double-stage THGEM setup using the Ar-CO$_2$ (90:10). For the latter case, the Ar-isobutane (95:5) gas mixture was used as well. 
The source was translated across the entire active area of the detector foil along X-direction, orthogonal to the strip length, over a total distance of $2.8\,\mathrm{cm}$ in 56 steps. A schematic representation of the procedure and the detector functioning in single-stage configuration is shown in figure ~\ref{fig:sourcemovement}. At each step, the source was displaced by $500\,\mathrm{\mu m}$ and kept at that position for $5\,\mathrm{mins}$ for data acquisition. The DAQ system and the positioning system stage were synchronized accordingly to ensure controlled movement of the source and data acquisition at fixed time intervals. Measurements were carried out for different voltage settings. To reduce the effect of foil charging-up, an interval of $24\,\mathrm{hrs}$ was maintained between successive voltage settings.  

\subsection{Analysis and results}
\begin{figure}[htbp]
\centering
   \subfloat[ \label{fig:strip_energy}]{%
    \includegraphics[width=0.49\textwidth]{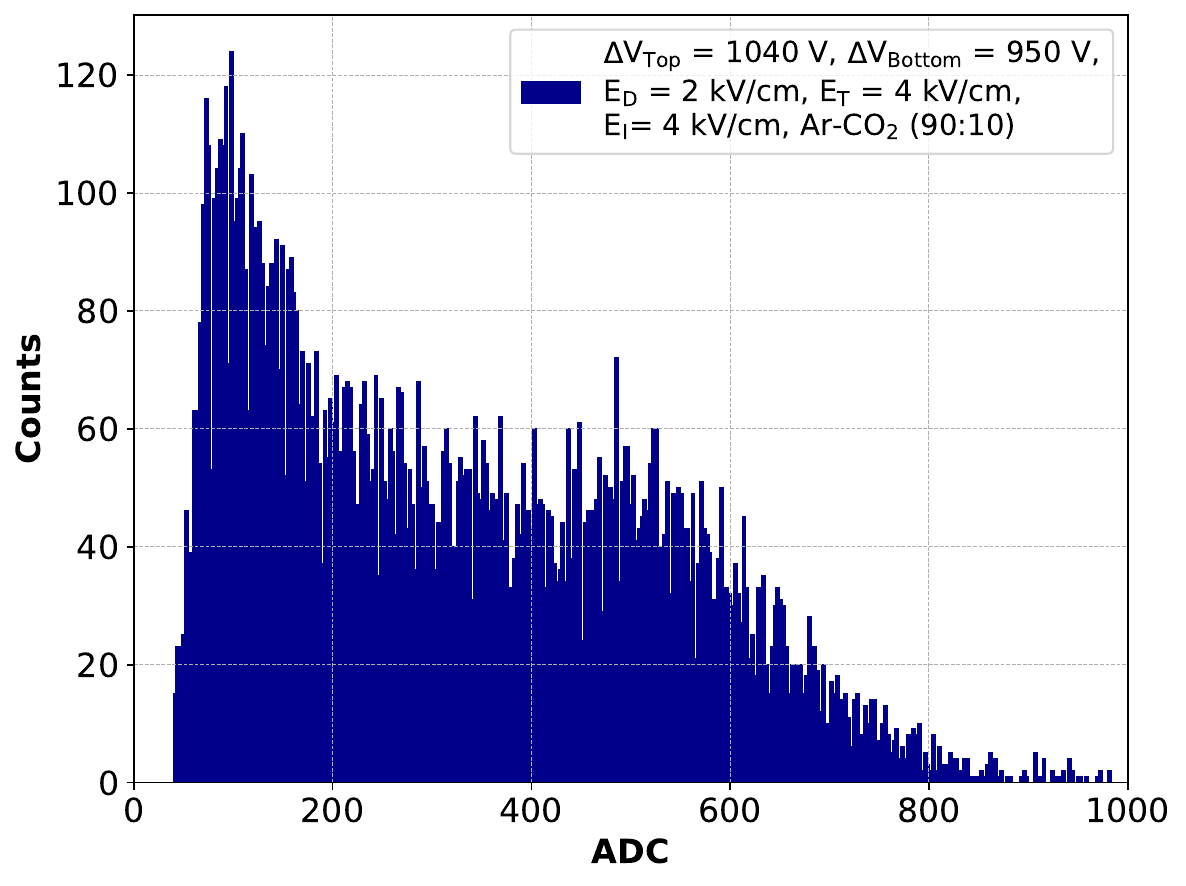}}
  \hfill
   \subfloat[\label{fig:event_spectrum}]{%
    \includegraphics[width=0.5\textwidth]{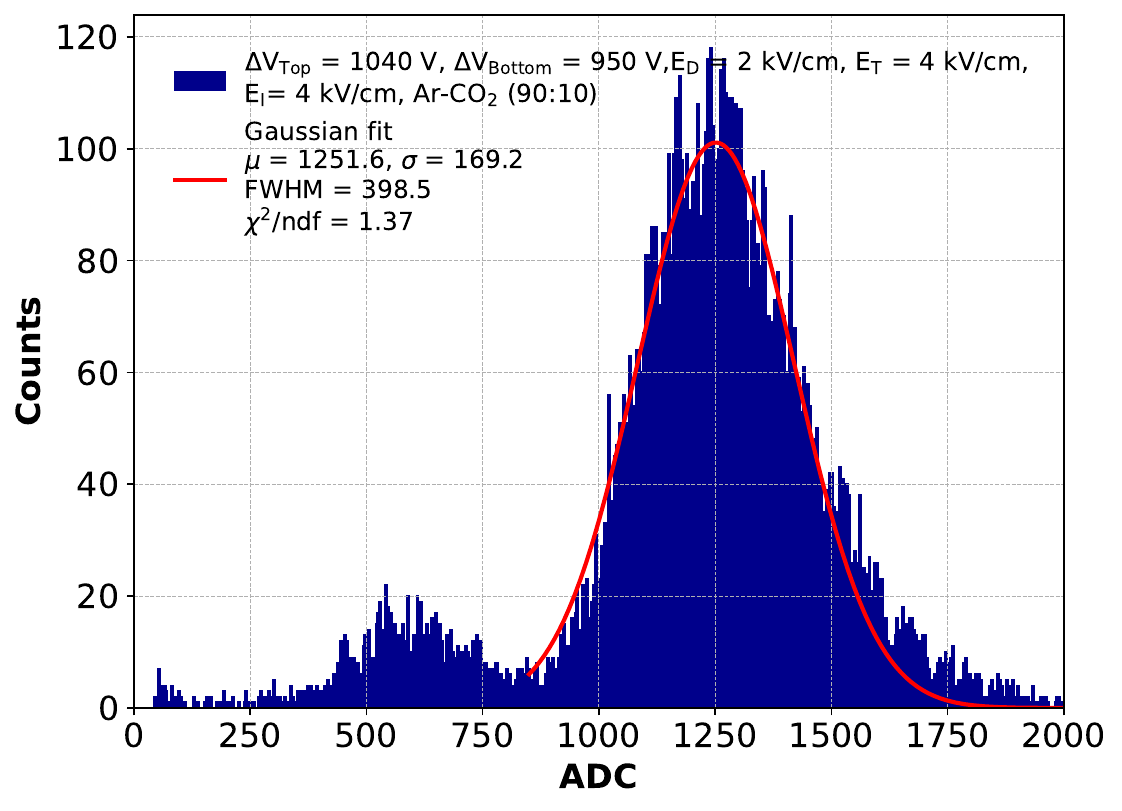}}
\caption[]{\label{fig:energy_spectrum_2} Distribution of charge deposit in terms of ADC counts (a) on a single readout strip, and (b) total charge deposition on consecutive readout strips over $100\,\mathrm{\micro s}$ time window}
\end{figure}

The data acquired with COMPASS software in CSV format were further analysed for event reconstruction and finally calculate the position resolution of the detector. 
For each X-ray event, the charge deposits on consecutive readout strips in terms of ADC counts were grouped using the corresponding time-tag information, recorded in the data-set, to determine the total charge collected on the readout for that event. The raw histogram of the ADC counts of a single readout strip is shown in figure~\ref{fig:strip_energy}, as an example, for double-stage operation using Ar-CO$_2$ (90:10). The pulse height spectrum obtained by adding the ADC values of the consecutive readout strips over a time window of $100\,\mathrm{\mu s}$ is presented in figure~\ref{fig:event_spectrum}. It shows the photo-peak corresponding to the $5.9\,\mathrm{keV}$ X-ray from Fe$^{55}$-source that was fitted with Gaussian function, as shown by the solid red line, to obtain the mean ADC count representing the charge deposit or the gas gain for the given voltage setting. 

\begin{figure}[htbp]
\centering
\includegraphics[width=.5\textwidth]{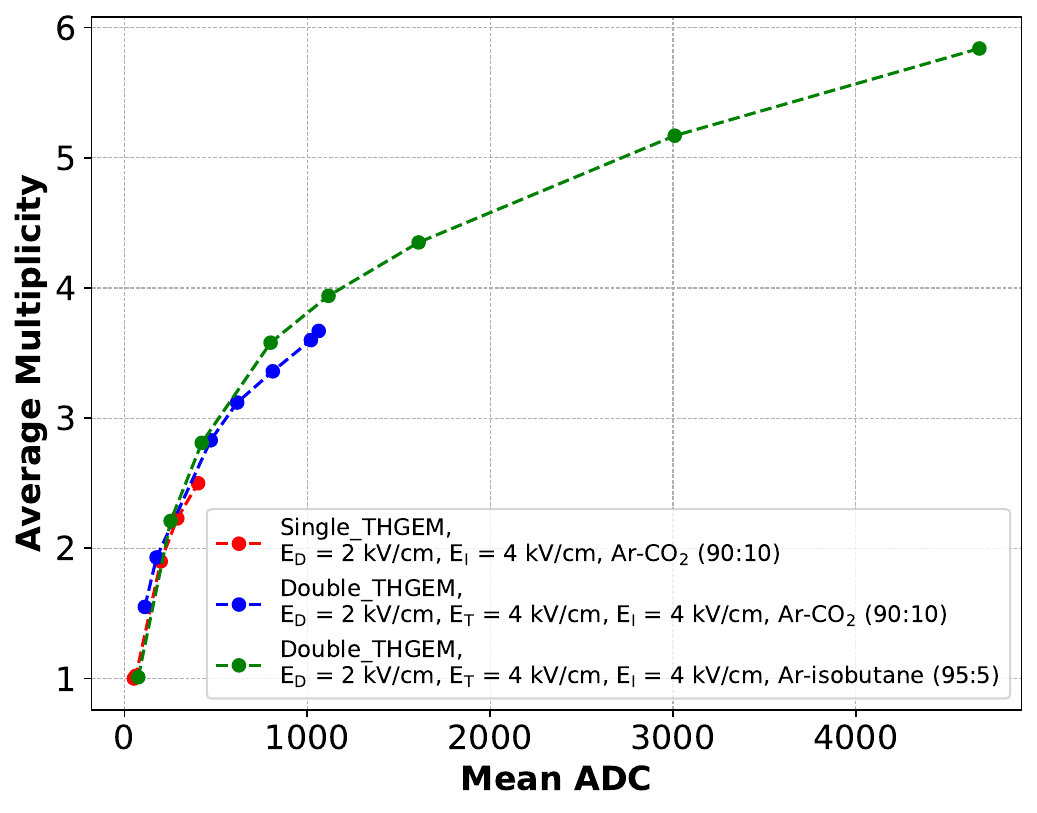}
\caption{Average strip multiplicity for a given source position with gain for single and double-stage configuration\label{fig:multiplicity_var}}
\end{figure}

The average hit multiplicity of the THGEM detector in single and double-stage operation using both the Argon-based gas mixture is plotted in figure~\ref{fig:multiplicity_var} as a function of the mean ADC count representing the detector gain. It shows that that the multiplicity increases with the gain and for the double-stage configuration, a fairly large multiplicity can be observed with Ar-isobutane (95:5) at higher gain.
 
For reconstruction of the position of the X-ray event from the recorded strip signals, the center-of-gravity (COG) method was followed. For this purpose, the data collected over an interval of $5\,\mathrm{mins}$ was considered which produced a distribution of counts over the strips, depending on the position of the source above the detector active area. 
 
\begin{figure}[htbp]
\centering
   \subfloat[ \label{fig:channel_count}]{%
    \includegraphics[width=0.5\textwidth]{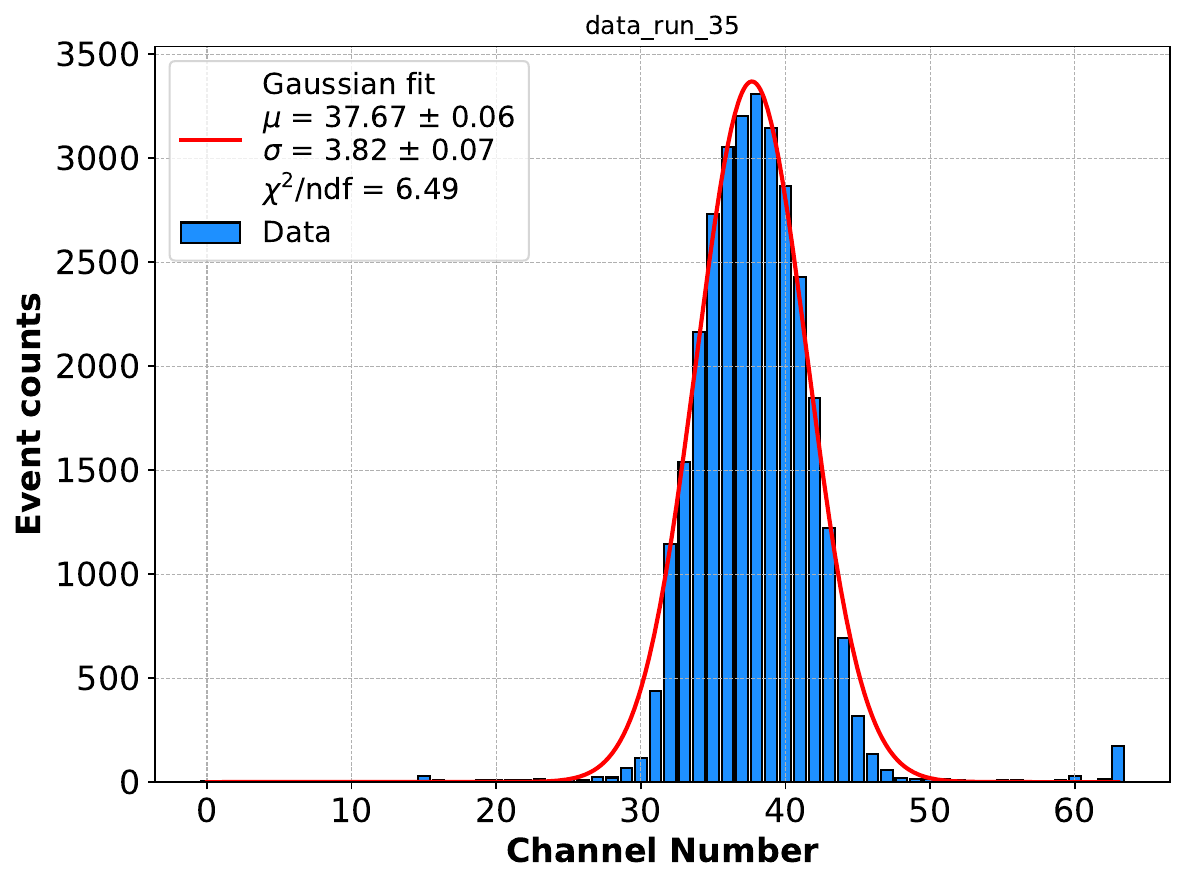}}
  \hfill
   \subfloat[\label{fig:position_calibration}]{%
    \includegraphics[width=0.5\textwidth]{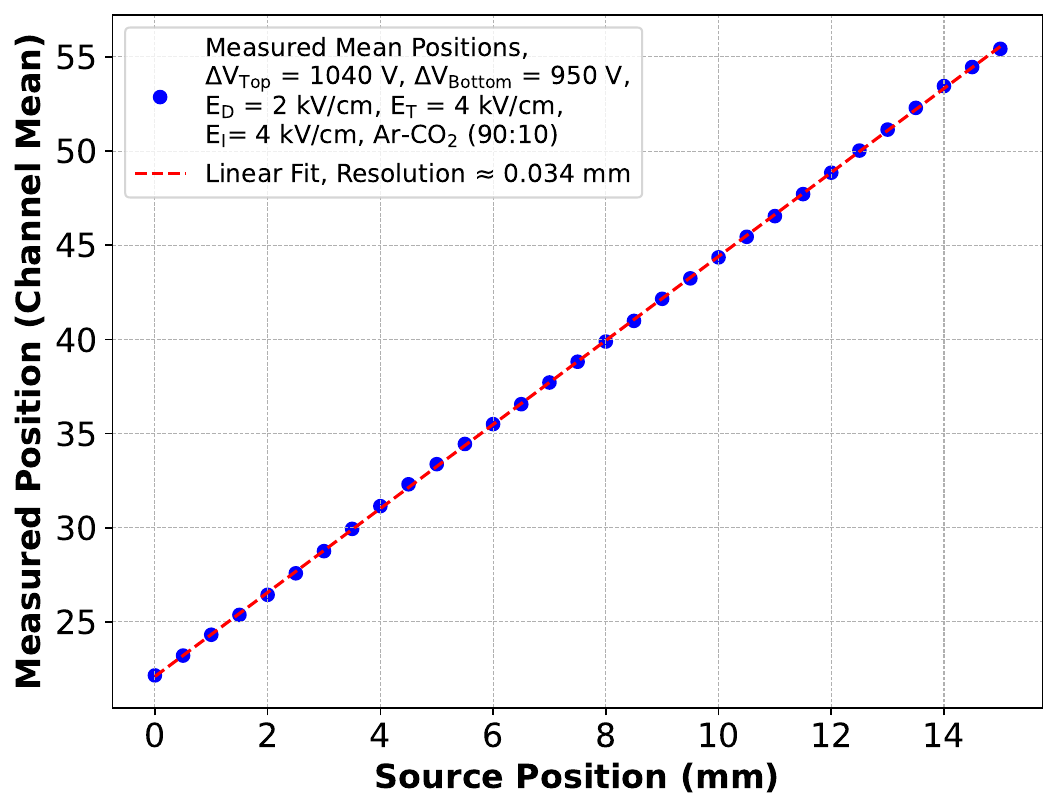}}
\caption[]{\label{fig:pos_calib} (a) Distribution of channel counts for run number 35  and (b) actual source position vs measured mean position }
\end{figure}

The strip-wise counts were therefore filled into a histogram containing all 64 channels. This distribution was fitted with a Gaussian function to determine its mean value, which was considered as the reconstructed source position. 
For a particular source position, the channel versus count distribution is shown in figure~\ref{fig:channel_count}, as an example. A total of 40 (N) reconstructed positions, determined from the recorded data, are plotted as a function of the corresponding actual source positions, provided by the positioning system, which followed a linear relationship as indicated by a linear fit, as shown in figure~\ref{fig:position_calibration} by the dashed red line. The residual of each position was defined as the difference between the measured position and the value predicted by the fitted line. The standard deviation of the residual gave us the position resolution of the detector. The error of the position resolution was determined following the formula given in equation~\ref{eqn:erroronposres}.
 \begin{equation}\label{eqn:erroronposres}
     \delta\sigma_{\text{residual}} \approx \sqrt{\frac{1}{2(N-1)}}\times\sigma_{\text{residual}}
 \end{equation}
 
\begin{figure}[htbp]
\centering
\includegraphics[width=.6\textwidth]{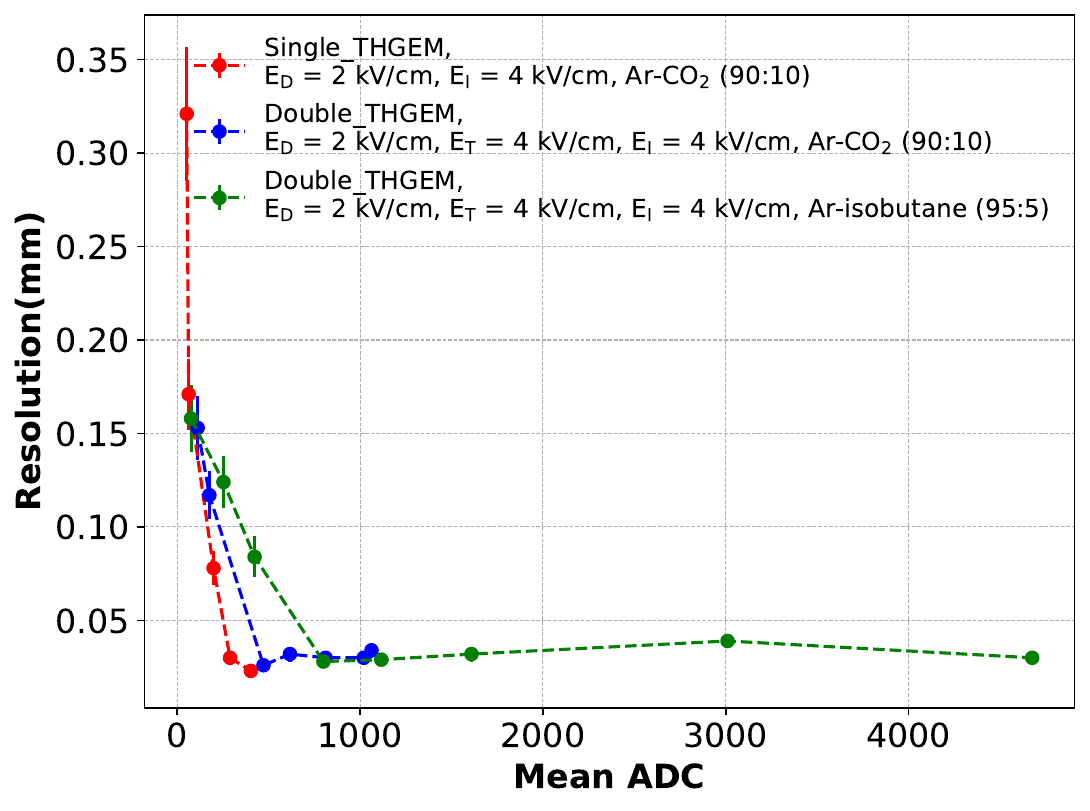}
\caption{Position resolution versus gain for single and double-stage THGEM detector using Ar-CO$_2$ (90:10) and Ar-isobutane (95:5)\label{fig:pos_res_gain_var}}
\end{figure}

Figure~\ref{fig:pos_res_gain_var} shows the measured position resolution for single and double-stage THGEM detectors. In case of single-stage THGEM operation using Ar–CO$_2$ (90:10), a clear improvement was observed with increase in the detector gain. Since a higher gain produces a larger number of avalanche electrons, resulting in charge sharing over multiple readout strips, it improves thereby the accuracy of the COG method to determine the position reconstruction. The best spatial resolution obtained for this case is $23\,\mathrm{\mu m}$ at the maximum gain. A similar trend was observed for double-stage operation in both Ar-CO$_2$ (90:10) and Ar-isobutane (95:5) gas mixtures. In both the gases, the resolution was poor at low gain-values, and improved as the gain increased, and eventually saturated at higher gain-values at around $30\,\mathrm{\mu m}$. 
It was also observed that in the low-gain region, for a given ADC value, the detector provided better spatial resolution in Ar-CO$_2$ (90:10) in comparison to that obtained using Ar-isobutane (95:5).
The excellent position resolution achieved in this work can be attributed to the high signal-to-noise ratio, with electronic noise remaining below 40 ADC channels, as well as effective charge sharing among multiple readout strips. The event positions were reconstructed using the Centre-of-Gravity (CoG) method, which enables sub-strip spatial interpolation and improves localization accuracy. As a result, a best position resolution corresponding to approximately 7–8$\%$ of the strip pitch was achieved, demonstrating the effectiveness of the detector and reconstruction technique~\cite{Hilke2020}.

\begin{figure}[htbp]
\centering
\includegraphics[width=.6\textwidth]{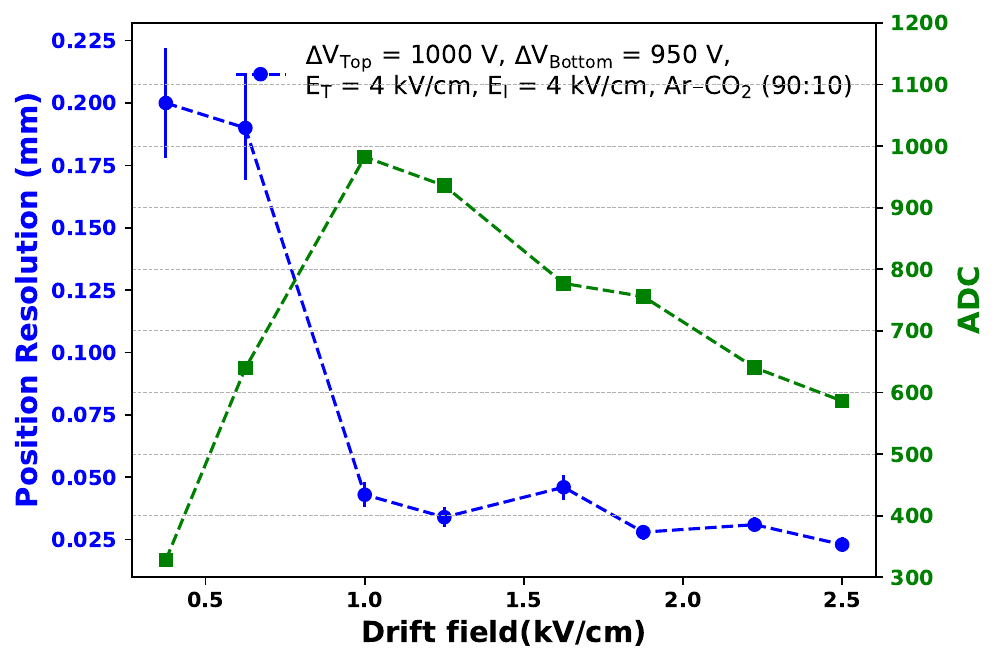}
\caption{Variation of position resolution with drift field.\label{fig:pos_res_drift_var}}
\end{figure}

Additional studies have been carried out with double THGEM configuration to understand the effect of drift field on the measured position resolution. 
At lower drift-field values, the measured spatial resolution is relatively poor because of the combined effects of lower effective detector gain, lower electron drift velocity, and larger transverse diffusion in the drift region. With increasing drift field, the effective gain improves, resulting in better signal formation and more reliable reconstruction of the interaction position. At the same time, the higher drift velocity reduces the time available for transverse diffusion, which further improves the spatial resolution. Although loss of electrons on the top surface of THGEM at higher drift fields leads to some reduction in detector gain, the spatial resolution remains largely unaffected as the combined effect of transverse diffusion and electron drift velocity compensate for this effect as shown in figure~\ref{fig:pos_res_drift_var}. The best achievable position resolution is found to be 25$\mu$m.

\section{Conclusion}
\label{Conclusion}
In this work, THGEM detectors have been developed, fabricated and characterized in collaboration with local industry in India as a step toward a THGEM-based muon imaging system. Prototype foils with different geometrical parameters were produced and subjected to systematic cleaning, polishing, and conditioning procedures in order to improve their high-voltage stability and reduce discharge probability. The conditioned detectors were then studied in detail in both single-layer and double-layer configurations using Ar-$\mathrm{CO}_2$ and Ar-isobutane based gas mixtures.

The gain characterization has shown that stable operation over a broad range of electric-field configurations can be achieved with these locally fabricated THGEM foils. In single-layer operation, the detector has exhibited a maximum gain beyond ~$6.0\times10^3$ in both the gas mixtures, while the measurement of gain at different drift and induction field values have helped identify the favorable and optimized operating region. In double-layer operation, significantly higher effective gain has been achieved, reaching a maximum value of about $3.3\times10^4$ in Ar--$\mathrm{CO}_2$ (90:10), while also benefiting from improved tolerance to discharges due to the distributed amplification across two stages.

For application in muon tomography, the response of detector to minimum-ionizing particles has been evaluated using a scintillator-based coincidence setup. The measured muon detection efficiency has been found to exceed 95\% over a substantial range of operating voltages for both single- and double-layer configurations, with a maximum efficiency of 99.5\%. These results demonstrate that the fabricated THGEM detectors provide sufficiently high efficiency for charged-particle tracking applications relevant for muon imaging.

The spatial resolution has been investigated using a 64-channel strip readout and a precision 3-axis positioning system with a collimated $^{55}$Fe source. Using a center-of-gravity reconstruction technique, the best spatial resolution obtained in single-layer operation has been $23~\mu$m, while double-layer operation has yielded values that saturate at around $30~\mu$m at higher gain. The measurements have also shown that the detector performance depends on gas composition and drift field, with Ar--$\mathrm{CO}_2$ providing better resolution than Ar--isobutane in the low-gain region, consistent with its lower transverse diffusion and higher drift-velocity. 

Overall, these results establish that the THGEM detectors fabricated locally in India are capable of achieving stable gas gain, high muon detection efficiency, and excellent spatial resolution required for muon imaging applications. The present study therefore validates the feasibility of THGEM detectors as a candidate tracker for muon tomography. Future work will focus on extending the fabrication to larger-area THGEMs, implementing two-dimensional readout, and integrating multiple tracking planes in order to build a full system for muon imaging studies.

\acknowledgments
The authors S. Ghosh, S. Dutta, N. Biswas and N. Majumdar are grateful to the Department of Atomic Energy (DAE), Govt. of India, and the Saha Institute of Nuclear Physics for providing necessary financial and infrastructural supports to accomplish the work. The author P. Roy acknowledges the financial support from the National Science Foundation (NSF) and the Department of Energy (DOE), U.S. The authors are thankful to Variable Energy Cyclotron Centre, Mrs. Lipy Das and Mr. Sudipta Barman for technical help and assistance. Finally, the authors sincerely thank the DRD1 Collaboration for valuable advices and discussions.

\appendix
\section{Appendix A}
The following table presents a comparison of PCB materials from two different manufacturers, ISOLA and ITEQ
\label{mat_table}
\begin{table}[htbp]
\centering
\caption{Comparison of PCB Materials}
\renewcommand{\arraystretch}{1.3}
\begin{tabular}{|p{4.5cm}|p{4cm}|p{4cm}|}
\hline
\textbf{Property} & \textbf{IS410} & \textbf{ITEQ IT-180A} \\
\hline

Electric Strength &
44 kV/mm &
45 kV/mm \\
\hline

Volume Resistivity (after moisture resistance) &
$5.0 \times 10^{8}$ M$\Omega\cdot$cm  &
$3.0 \times 10^{10}$ M$\Omega\cdot$cm \\
\hline

Surface Resistivity (after moisture resistance) &
$8.0 \times 10^{6}$ M$\Omega$ &
$3.0 \times 10^{10}$ M$\Omega$ \\
\hline

Dielectric Breakdown &
$>50$ kV &
60 kV \\
\hline

Coefficient of Thermal Expansion (CTE) &
X/Y: 11 ppm/$^\circ$C \newline
Z: 55 ppm/$^\circ$C &
X/Y: 10--13 ppm/$^\circ$C \newline
Z($\alpha_1$): 45 ppm/$^\circ$C \newline
Z($\alpha_2$): 210 ppm/$^\circ$C \\
\hline

Permittivity (Dk) &
3.96 at 100 MHz \newline
3.90 at 1 GHz \newline
3.87 at 5 GHz \newline
3.87 at 10 GHz &
4.4 at 1 GHz \newline
4.3 at 2 GHz \newline
4.1 at 5 GHz \newline
4.1 at 10 GHz \\
\hline

Glass Transition Temp. (Tg) &
$180^\circ$C &
$180^\circ$C \\
\hline

Moisture Absorption &
0.20\% &
0.12\% \\
\hline

Flexural Strength &
82,600 lb/in$^2$ (lengthwise) &
84,300 lb/in$^2$ (lengthwise) \\
\hline

Max. Operating Temp. &
$130^\circ$C &
$130^\circ$C \\
\hline

\end{tabular}
\label{tab:pcb_materials}
\end{table}

\end{document}